\documentclass[11pt]{article}
\usepackage{comment,soul}
\usepackage[hidelinks]{hyperref}
\usepackage{enumitem}
\usepackage{amsmath,amssymb,epsf,cite,graphicx,subfigure}
\usepackage{amsmath,braket,tikzsymbols}
\usepackage[bbgreekl]{mathbbol}
\usepackage{comment}
\usepackage[export]{adjustbox}
\usepackage{dsfont}
\usepackage{faktor}
\usepackage{mathrsfs}
\usepackage{float}

\numberwithin{equation}{section}

\definecolor{airforceblue}{rgb}{0.36, 0.54, 0.66}
\newcommand{\beq}{\begin{equation}}
\newcommand{\eeq}{\end{equation}}

\begin{document}
\baselineskip=15.5pt
\pagestyle{plain}
\setcounter{page}{1}

\begin{center}
{\LARGE \bf Non-perturbative de Sitter Jackiw-Teitelboim gravity}
\vskip 1cm

\textbf{Jordan Cotler$^{1,a}$ and Kristan Jensen$^{2,b}$}

\vspace{0.5cm}

{\it ${}^1$ Society of Fellows, Harvard University, Cambridge, MA 02138, USA \\}
{\it ${}^2$ Department of Physics and Astronomy, University of Victoria, Victoria, BC V8W 3P6, Canada\\}

\vspace{0.3cm}

{\tt  ${}^a$jcotler@fas.harvard.edu, ${}^b$kristanj@uvic.ca\\}

\medskip

\end{center}

\vskip1cm

\begin{center}
{\bf Abstract}
\end{center}
\hspace{.3cm} 
With non-perturbative de Sitter gravity and holography in mind, we deduce the genus expansion of de Sitter Jackiw-Teitelboim (dS JT) gravity.  We find that this simple model of quantum cosmology has an effective string coupling which is pure imaginary. This imaginary coupling gives rise to alternating signs in the genus expansion of the dS JT $S$-matrix, which as a result appears to be Borel--Le Roy resummable.  Furthermore dS JT gravity is formally an analytic continuation of AdS JT gravity, and behaves like a matrix integral with a negative number of degrees of freedom.

\newpage

\tableofcontents

\section{Introduction}

Our contemporary understanding of cosmology is that we live in an expanding universe with a positive cosmological constant, which will tend towards de Sitter space in the far future~\cite{peebles2003cosmological}.  Even though there has been significant progress in understanding non-perturbative aspects of quantum gravity with a negative cosmological constant~\cite{Maldacena:1997re, Witten:1998qj}, the case of a positive cosmological constant -- most pertinent to our own universe -- remains shrouded in mystery.  Recently there has been progress on de Sitter quantum gravity in low-dimensional toy models, most notably nearly-dS$_2$ Jackiw-Teitelboim (dS JT) gravity~\cite{Maldacena:2019cbz, Cotler:2019nbi, Cotler:2019dcj, Cotler:2023eza} (see also~\cite{Anninos:2018svg}).  This model has the virtue of being analytically tractable while being sufficiently non-trivial that it may enable a broader understanding of de Sitter quantum gravity.

dS JT gravity has a better-known cousin, a negative cosmological constant version, which has a computable sum over topologies and a known holographic dual, a certain formal matrix integral~\cite{Saad:2019lba}. That matrix integral defines the bulk non-perturbatively. The situation in dS JT gravity is less understood. Some features of its putative genus expansion have been mapped out in~\cite{Maldacena:2019cbz, Cotler:2019nbi, Cotler:2019dcj}, but so far only three cases are understood: a divergent sphere partition function, the leading contribution to the wavefunction of a no-boundary state where the spacetime is a topological disk, and a Big Bounce cosmology~\cite{Cotler:2019nbi, Cotler:2019dcj,Cotler:2023eza} where the spacetime is a Lorentzian cylinder. The basic difficulty is that dS JT gravity includes a sum over spacetimes with constant positive scalar curvature, and it is not clear what spaces to sum over beyond the disk and cylinder.

We resolve that problem in this paper. We find that there is a well-defined topological expansion, which requires a careful treatment of the path integral measure together with an $i\epsilon$ prescription. The requisite spaces have complex time contours in which the geometry goes from (i) being Lorentzian in the far past, to (ii) having $(-,-)$ signature at intermediate times, to (iii) being Lorentzian in the far future. Equivalently, the spacetime can have $(-,-)$ signature throughout, as suggested by~\cite{Maldacena:2019cbz} for the Hartle-Hawking state. We choose a path integral measure that leads to a positive norm on states in the Lorentzian continuation.  Our analysis reveals a pure imaginary effective string coupling for the genus expansion. Moreover, we establish that the model's $S$-matrix obeys topological recursion.

Since the AdS version of JT gravity is dual to a formal matrix integral it is natural to ask if the same is true of the dS version. We find that dS JT gravity behaves like a formal matrix integral with a negative number of degrees of freedom. For instance, while an $N \times N$ matrix has $N^2$ entries, a de Sitter analogue would have $-N^2$ entries.  Since for a finite matrix model the number of effective degrees of freedom is $N_{\text{eff}} = N^2$, it is as if we are taking $N_{\text{eff}} \to - N_{\text{eff}}$, which is reminiscent of work in de Sitter Vasiliev gravity~\cite{Anninos:2011ui}.\footnote{There have been other suggestions from other points of view that the static patch of de Sitter has a dual with a negative number of degrees of freedom. See e.g.~\cite{Anninos:2022ujl} for a discussion.} While it may be tempting to try to posit a dual with fermionic degrees of freedom, we later describe some obstructions to this possibility.

In the remainder of the paper we review dS JT gravity and then establish its topological expansion.  In so doing we clarify the relation of the model to Euclidean AdS JT gravity.  We then turn to some consequences of the topological expansion, study non-perturbative effects, and conclude with a discussion. In the Appendix we give additional details about a de Sitter analogue of the Airy model, and also discuss the dS JT Klein-Gordon inner product of~\cite{Maldacena:2019cbz}.

\section{Preliminaries}

\subsection{Overview of dS JT gravity}

We begin by reviewing the basic ingredients of dS JT gravity~\cite{Maldacena:2019cbz, Cotler:2019nbi}.  The action of the theory is
\begin{equation}
\label{E:action1}
	S_{\text{JT}} = \frac{S_0}{4\pi} \int_M d^2 x \sqrt{-g}\,R + \int_M d^2 x \sqrt{-g} \,\phi(R - 2) + S_{\text{bdy}}\,,
\end{equation}
where $S_0\gg 1$, suppressing fluctuations of topology, and
\begin{align}
\label{E:action2}
	S_{\text{bdy}} = - \frac{S_0}{2\pi} \int_{\partial M}  dx \sqrt{\gamma}\,K - 2 \int_{\partial M} dx \sqrt{\gamma}\,\phi(K-1)\,,
\end{align}
where here $\gamma$ is the induced metric on the boundary and $K$ is the corresponding extrinsic curvature. 
The standard boundary conditions for the metric and dilaton near future infinity $t\to\infty$ are
\begin{equation}
\label{E:bcs1}
ds^2 = -dt^2 + (e^{2t} + O(1))dx^2\,,\quad \phi = \frac{\Phi}{2\pi}\,e^{t} + O(1)\,,
\end{equation}
where $x \sim x + 2\pi$ and $\Phi$ is a real constant with a small positive imaginary part.\footnote{Although one can allow for more general, $x$-dependent $\Phi$'s as in~\cite{Cotler:2023eza}, this does not affect our results in this paper.}  There are similar boundary conditions for $t \to - \infty$. Crucially, the physics of the model only depends on the ratio of the renormalized dilaton (here $\frac{\Phi}{2\pi}$) to the renormalized length of the boundary circle (here $2\pi$)~\cite{Cotler:2019nbi}, and so the asymptotic states are superpositions of $|\Phi\rangle$ where $\Phi \in \mathbb{R} + i \epsilon$ in the far future and $\Phi \in \mathbb{R} - i \epsilon$  in the far past~\cite{Cotler:2019dcj}.  As we discuss in Section~\ref{subsub:cyl}, the $i\epsilon$ prescription is required so that the moduli space integral that computes the global de Sitter amplitude converges.  We also note that classical dS JT gravity has a time-reversal symmetry which exchanges past and future boundary conditions.

\subsection{What are we summing over?}

The basic object computed by dS JT gravity is the $S$-matrix. In the infinite past we may prepare an in-state $|\text{in}\rangle = |\Phi_1',...,\Phi_{n_{\rm P}}'\rangle$ composed of $n_{\rm P}$ asymptotically large circles where each has some value of $\Phi'$, and similarly in the far future we consider an out-state $\langle \text{out}| = \langle \Phi_1,...,\Phi_{n_{\rm F}}|$ composed of $n_{\rm F}$ asymptotically large circle which each has some value of $\Phi$. The JT path integral over surfaces which fill in these boundary conditions computes an unnormalized $S$-matrix element 
\vspace{6pt}
\begin{align}
\includegraphics[scale=.37, valign = c]{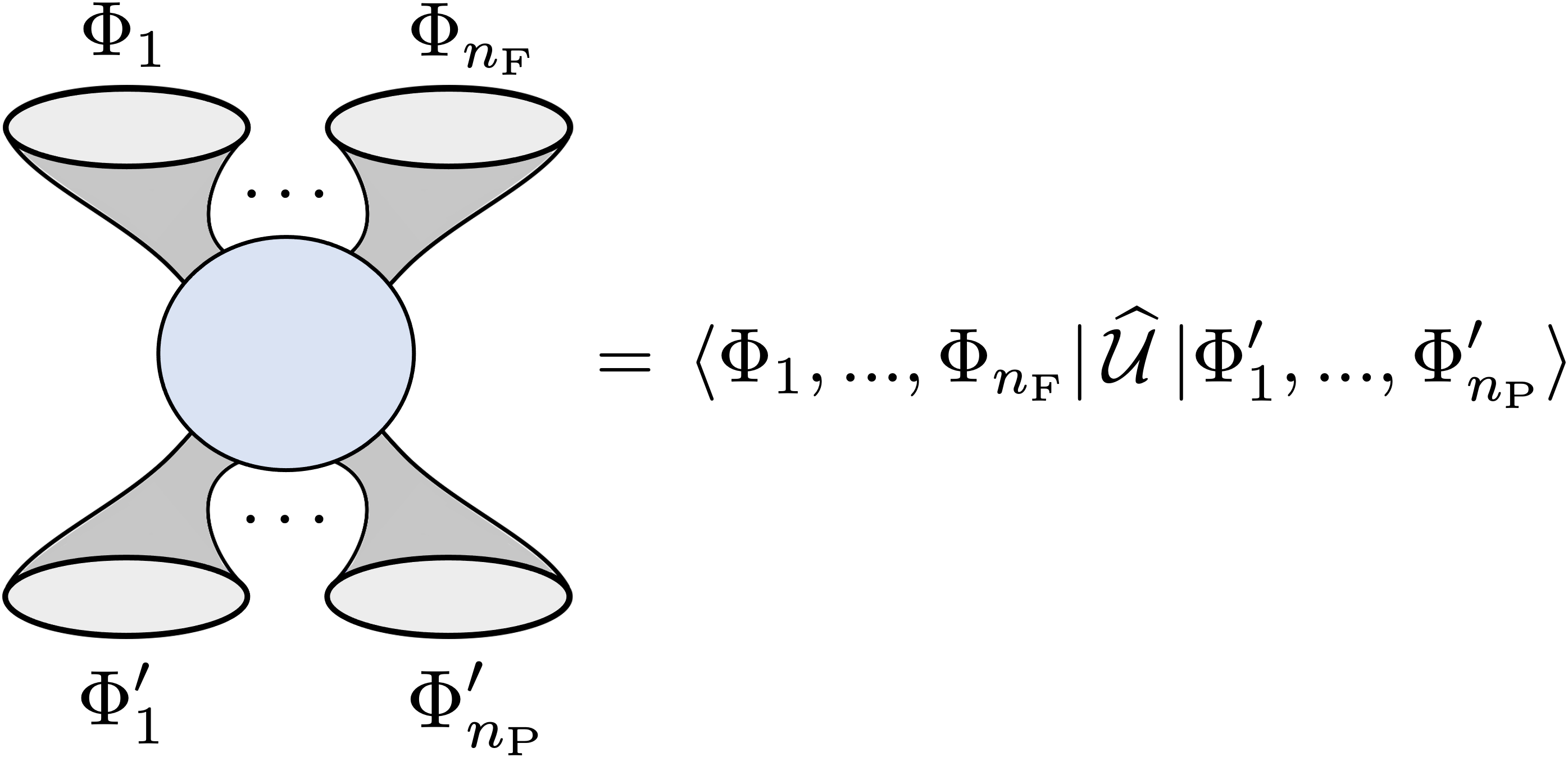}
\end{align}
where $\widehat{\mathcal{U}}$ denotes the infinite-time evolution operator. The left-hand-side refers to the sum over all surfaces that fill in the boundary conditions specified by the initial and final states weighted by $e^{iS_{\text{JT}}}$. In principle we may consider $n_{\rm P}=0$ and/or $n_{\rm F}=0$, corresponding to a no-boundary initial and/or final state.

\begin{figure}[t!]
\begin{center}
\includegraphics[width = \textwidth]{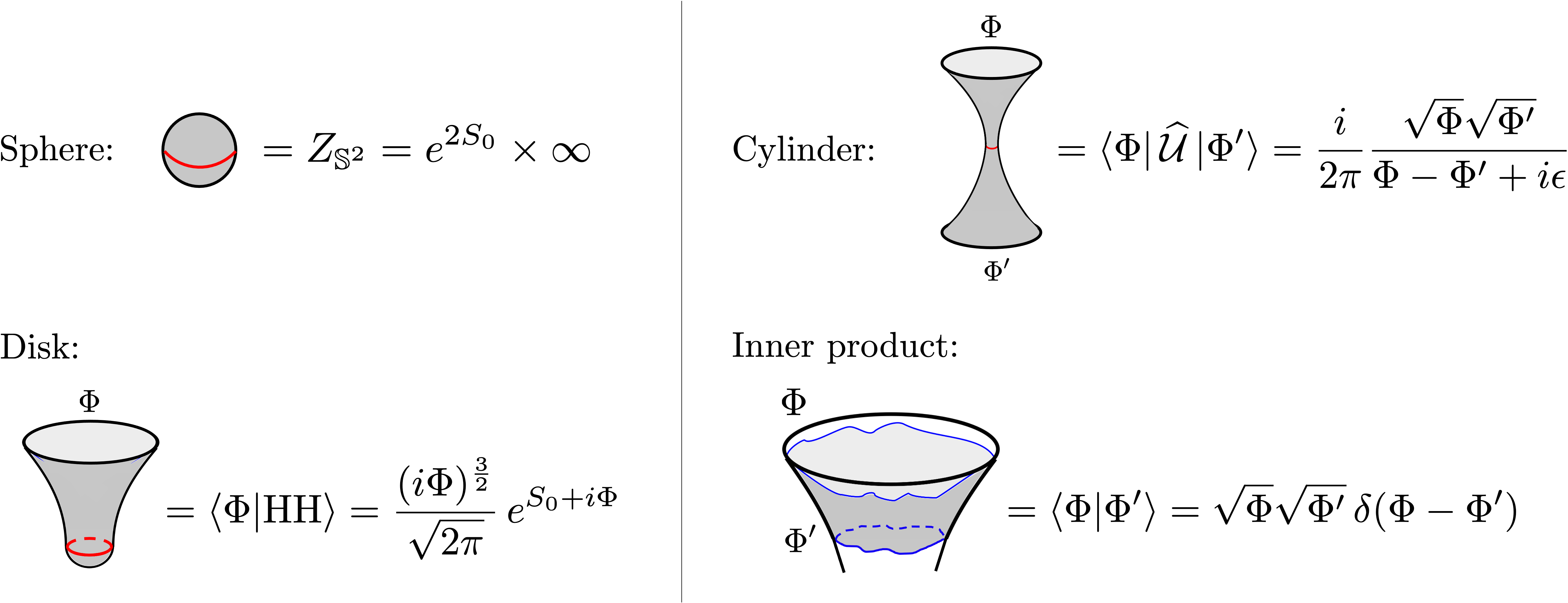}
\vspace{.1cm}
\caption{\label{fig:amps1} A depiction of the sphere, disk (Hartle-Hawking), and cylinder (global dS$_2$) amplitudes, as well as the inner product on single-boundary asymptotic states.}
\end{center}
\end{figure}

In Euclidean AdS JT gravity the analogous Euclidean amplitude has a genus expansion in powers of an effective string coupling $\exp(-S_0)$. What about the dS version?  There is evidence that a similar statement holds in dS JT gravity. Three amplitudes of dS JT gravity are known, with the spacetime being either (i) a sphere (which has no asymptotic circles), (ii) a disk (Hartle-Hawking state), which has one past or future asymptotic circle~\cite{Maldacena:2019cbz, Cotler:2019nbi}, or (iii) a cylinder (global dS$_2$), which connects a past asymptotic circle to a future asymptotic circle~\cite{Cotler:2019nbi}.  There is also a fourth known quantity which is not an amplitude, but rather the inner product on single-boundary asymptotic states~\cite{Cotler:2019dcj}.  The amplitudes and inner product are depicted in Fig.~\ref{fig:amps1}.  The phases in front of these amplitudes depend on the details of the path integral measure, which we carefully address in the next Section.

The sphere amplitude is proportional to $\exp(2S_0)$ but is one-loop divergent. Since the disk has only one asymptotic boundary which we can take to be in the far future, the disk amplitude computes the leading approximation to the Hartle-Hawking wavefunction of the no-boundary state. The corresponding two-dimensional spacetime is given by the metric
\begin{align}
\begin{cases}
ds^2 = -dt^2 + \cosh^2(t)\,dx^2 & \text{for } t\geq 0 \\
ds^2 = d\tau^2 + \cos^2(\tau)\,dx^2 & \text{for } \tau\in (-\pi/2,0]
\end{cases}\,,
\end{align}
which is the union of a Lorentzian segment for $t \geq 0$ comprising half of global dS$_2$, and a Euclidean hemisphere for $\tau \in (-\pi/2,0]$ which caps off the spacetime and prepares the no-boundary state.  The Lorentzian and Euclidean parts are smoothly glued together with the $t=0$ circle glued to the equator at $\tau=0$.  The term proportional to $S_0$ in the dS JT action $S_{\rm JT}$ evaluates to $-iS_0$ so that the disk amplitude $\sim e^{iS_{\rm JT}}$ is weighted by a factor of $\exp(S_0)$. While the disk requires a Euclidean segment, the cylinder does not, and so for the cylinder one can sum over smooth Lorentzian metrics. The cylinder amplitude is proportional to $\exp(0\times S_0)$.

The norm on single-boundary asymptotic states $\langle \Phi | \Phi'\rangle = \sqrt{\Phi} \sqrt{\Phi'}\,\delta(\Phi - \Phi')$ was computed in~\cite{Cotler:2019dcj}, which enabled an analysis of the unitarity (or lack thereof) of the $S$-matrix.  This analysis was carried out in~\cite{Cotler:2023eza} for the cylinder amplitude, revealing that infinite-time evolution in global dS$_2$ is a composition of a co-isometry and an isometry.

Aside from the aforementioned three amplitudes and the inner product on single-boundary states, the general picture is unclear. Integrating out the dilaton enforces the constant curvature condition $R=2$, and it is easy to prove, on a general surface, that there are no smooth $R=2$ metrics of purely Lorentzian or Euclidean signature over which to sum~\cite{Cotler:2019nbi}. So it is not immediately obvious how to treat other topologies.

\begin{figure}[t]
\begin{center}
\includegraphics[width = .8\textwidth]{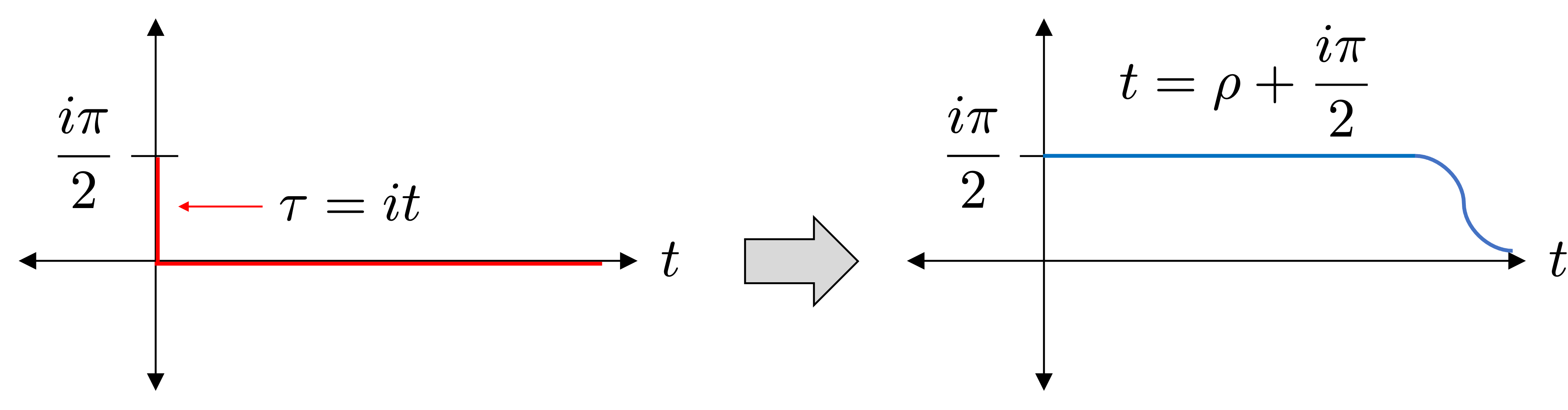}
\caption{\label{fig:tcontour} Two different complex time contours for the dS JT Hartle-Hawing geometry. The red contour corresponds to half of global dS$_2$, glued to half of a Euclidean hemisphere, while the metric on the blue contour  is that of the hyperbolic disk in $(-,-)$ signature. Both contours connect the endpoints $t=i\pi/2$ and $t\to\infty$.}\end{center}
\end{figure}

The disk amplitude mentioned above offers a path forward. Let us write the spacetime line element and its corresponding dilaton profile as
\begin{align}
\begin{cases}
\label{E:HHfuture}
ds^2 = -dt^2 + \cosh^2(t)\,dx^2 \\
\,\,\,\,\,\phi = \frac{\Phi}{2\pi}\,\sinh(t)
\end{cases}\,,
\end{align}
where we interpret $t$ as following a complex time contour connecting $t=i\pi/2$ to $t\to\infty$ (see the red curve in Fig.~\ref{fig:tcontour}). There is another contour~\cite{Maldacena:2002vr, Maldacena:2011mk} that connects those two points, given by $t=\rho + \frac{i \pi}{2}$ with $\rho\geq 0$ (see the blue curve in Fig.~\ref{fig:tcontour}). The line element and corresponding dilaton profile along that contour is 
\begin{align}
\label{E:continuation1}
\begin{cases}
ds^2 = -\left(  d\rho^2 +\sinh^2(\rho)\,dx^2\right)\\
\,\,\,\,\,\phi = \frac{i\Phi}{2\pi}\,\cosh(\rho)
\end{cases}\,.
\end{align}
The metric is nothing more than the hyperbolic disk in $(-,-)$ signature, which thanks to the unconventional signature has $R=2$. More generally, we can take any complex time contour that connects the Euclidean cap at $t=i\frac{\pi}{2}$ to the nearly dS$_2$ region as $t\to\infty$.

In fact, \emph{any} smooth hyperbolic metric in $(-,-)$ signature has $R=2$. This gives us our desired, candidate set of metrics on a general surface, namely hyperbolic metrics in $(-,-)$ signature.  Indeed the authors of~\cite{Maldacena:2019cbz} suggested exactly this in the context of the genus expansion of the Hartle-Hawking wavefunction. Moreover,~\eqref{E:continuation1} suggests that the hyperbolic $(-,-)$ signature metrics should be equipped with an imaginary dilaton profile.

Now we should pause to check that these candidate metrics are consistent with our boundary conditions. Recall that asymptotic boundary conditions in dS JT gravity are labeled by a single real number $\Phi$ for each boundary circle. When that circle is in the far future, it is just the ratio of the renormalized dilaton to the renormalized length of the boundary circle.\footnote{This parameter appears in the JT path integral in the following way. Each asymptotic circle carries an independently fluctuation Schwarzian mode described by an action with two coupling constants. One is precisely this parameter $\Phi$.} For a hyperbolic metric in $(-,-)$ signature the renormalized length is now imaginary, which means that we can retain the same boundary condition as in Lorentzian signature provided that the renormalized dilaton is also imaginary. In an equation, the boundary condition for a future circle characterized by $\Phi$ now reads:
\beq
\label{E:minusBCs}
\text{Future b.c.'s:} \qquad \begin{cases}
ds^2 = -\left( d\rho^2 + (e^{2\rho}+O(1))\,dx^2\right) \\
\,\,\,\,\,\phi = \frac{i \Phi}{2\pi}\,e^{\rho}  + O(1)
\end{cases}\,,\quad \text{as }\rho \to \infty\,.
\eeq
We observe that these boundary conditions are indeed consistent with~\eqref{E:continuation1}.  For completeness, the boundary condition for a past circle characterized by $\Phi$ is 
\beq
\label{E:minusBCs2}
\text{Past b.c.'s:} \qquad \begin{cases}
ds^2 = -\left( d\rho^2 + (e^{-2\rho}+O(1))\,dx^2\right) \\
\,\,\,\,\,\phi = -\frac{i \Phi}{2\pi}\,e^{-\rho}  + O(1)
\end{cases}\,,\quad \text{as }\rho \to -\infty\,.
\eeq
\begin{figure}[t!]
\begin{center}
\includegraphics[width = .3\textwidth]{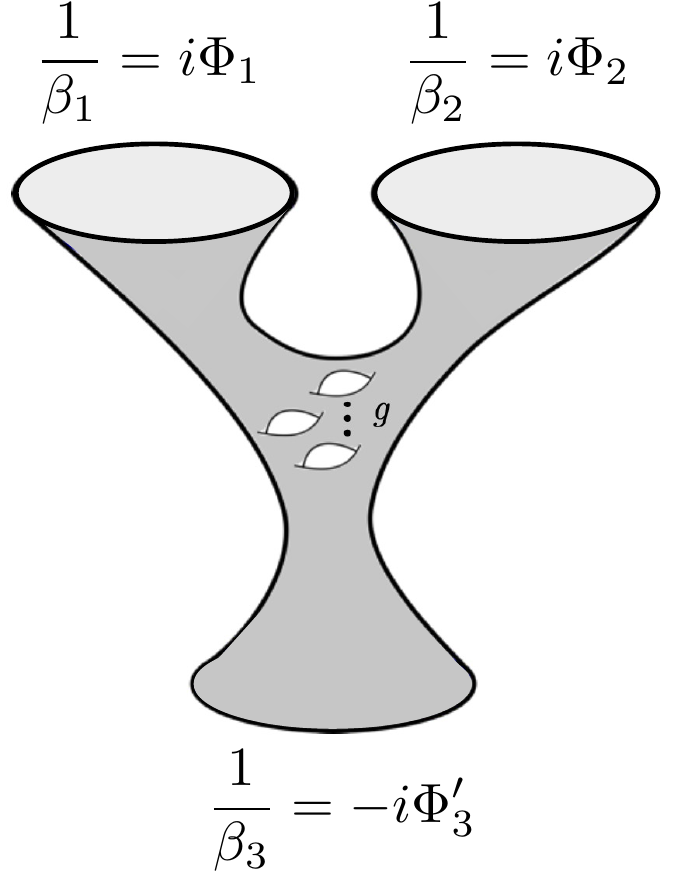}
\vspace{.1cm}
\caption{\label{fig:minusminus1} A schematic of a higher-genus geometry included in the dS JT path integral.  It is described by a hyperbolic metric in $(-,-)$ signature, which then has $R = 2$, and we take the boundary `inverse temperatures' $\beta$ to be pure imaginary with infinitesimally negative real part, as required by the the boundary conditions of dS JT gravity.}
\end{center}
\end{figure}
\noindent These boundary conditions are analytic as a function of complex time at infinity. As a result, imposing the desired Lorentzian boundary conditions~\eqref{E:bcs1} near future infinity is equivalent to imposing~\eqref{E:minusBCs}, while~\eqref{E:minusBCs2} is equivalent to the Lorentzian boundary condition near past infinity.

\begin{figure}[h!]
\begin{center}
\includegraphics[width=.4 \textwidth]{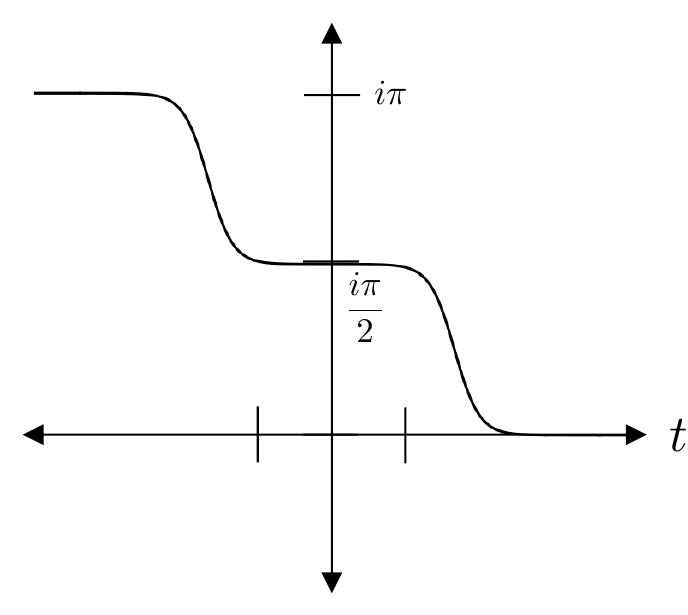}
\caption{\label{F:minusCTC} A complex time contour for the geometry in Fig.~\ref{fig:minusminus1} so that the geometry is asymptotically Lorentzian but has a $(-,-)$ signature region at intermediate times.}
\end{center}
\end{figure}

Our proposal is then to define dS JT amplitudes on a genus $g$ surface with $n$ boundaries through a sum over hyperbolic metrics in $(-,-)$ signature with the boundary conditions in~\eqref{E:minusBCs} and~\eqref{E:minusBCs2}. Equivalently, we can consider geometries with a complex time contour, where the geometry is Lorentzian near infinity while interpolating through a region in which it has $(-,-)$ signature. A schematic of these spacetimes is depicted in Fig.~\ref{fig:minusminus1}, and a complex time contour for it in Fig.~\ref{F:minusCTC}. Our proposal implies a genus expansion, since the term proportional to $S_0$ in the action guarantees that the amplitude on a genus $g$ surface with $n$ boundaries is proportional to $\exp\!\left( \chi S_0\right)$ where $\chi=2-2g-n$ is the Euler characteristic. So far, our discussion suggests that dS JT gravity is an analytic continuation of its Euclidean AdS cousin (in $(-,-)$ signature), where the renormalized dilaton is rotated from positive real values to imaginary ones (a viewpoint taken in~\cite{Cotler:2019nbi}). However this is not the full story. One hint that the story is more complicated is that taking this prescription seriously leads to a negative-definite norm on states.  In fact, the correct non-perturbative theory of dS JT gravity requires a careful treatment of the phase in front of the path integral measure and the implementation of an $i \epsilon$ prescription.  We explain these ingredients in detail in the next Section.

\section{Genus expansion}

The goal of this Section is to compute the amplitudes of dS JT on a genus $g$ surface with $n$ boundaries.  A crucial step involves adopting an appropriate measure on the space of constant curvature metrics, which is related to but distinct from the measure arising in the Euclidean AdS setting. The appropriate measure leads to sensible results, including a positive norm on states in global dS.  Our end result is detailed in~\eqref{E:masterAmplitude} and~\eqref{E:masterAmplitude2}, and can be understood as a particular analytic continuation of the amplitudes of Euclidean AdS JT gravity. See the end of this Section for a summary of the main results.

To elaborate a bit further, note that after integrating out the dilaton we have a residual integral to perform over the moduli space of hyperbolic metrics in $(-,-)$ signature. This moduli space is well-understood. Let us take a surface $\Sigma$ of genus $g$ with $n$ asymptotic boundaries.\footnote{The disk and cylinder have to be treated separately.} $\Sigma$ can be decomposed by a cut-and-paste procedure into an intermediate genus $g$ surface $\Sigma_0$ with $n$ geodesic borders of lengths $b_1,b_2,\hdots,b_n$, and $n$ external ``trumpets'' glued to those geodesic borders, connecting them to the $n$ asymptotic boundaries. The total moduli space is symplectic with a sympletic form $\Omega$ and one can assign coordinates on it in the following way. There are $3g+n-3$ pairs of moduli associated with $\Sigma_0$ that make up the Weil-Petersson moduli space; there are another $n$ pairs of moduli associated with how the trumpets are glued to $\Sigma_0$ (these pairs are composed of the $b_i$ and their symplectic partners); finally, associated to each trumpet there is a reparameterization of its asymptotic circle (the Schwarzian mode on its boundary) -- an element of $\faktor{\text{Diff}(\mathbb{S}^1)}{U(1)}$ -- which is a coadjoint orbit of the Virasoro group and thus symplectic. Because the residual integration space is symplectic, there is a natural positive-definite volume form on it, namely $\text{Pf}(\Omega)$. Euclidean AdS JT amplitudes are obtained by integrating with respect to this measure. 

We claim that the right measure for dS JT gravity is
\beq
\label{E:theMeasure}
	\text{Pf}(-\Omega)\,,
\eeq
together with a 90$^{\circ}$ rotation of the contour of integration for the Schwarzian modes mandated by our $i\epsilon$ prescription. We will justify this claim shortly. First let us explain the proposal in more detail. For a finite-dimensional symplectic space of dimension $2k$, we have $\text{Pf}(-\Omega) = (-1)^k \text{Pf}(\Omega)$. Our moduli spaces are infinite-dimensional thanks to the Schwarzian modes and require a bit more care which we undertake presently.

\subsection{Disk and trumpet}

Let us begin with the Schwarzian path integral describing the disk with a boundary at future infinity, or past infinity, corresponding to the Hartle-Hawking wavefunction of the no-boundary state in the far future or far past, respectively. There is a Schwarzian mode living on the asymptotic circle. Using the measure $\text{Pf}(-\Omega)$, the disk amplitude of dS JT gravity reads\footnote{We use a tilde to distinguish the dS JT amplitudes from AdS JT amplitudes.}
\begin{align}
\label{E:diskpathint1}
\widetilde{Z}_{\rm disk}(\beta) = e^{S_0}\int  \text{Pf}(-\Omega)\,e^{\frac{1}{\pi \beta} \int_0^{2\pi} du\, \left( \{f(u),u\} + \frac{1}{2}f'(u)^2\right)}\,,
\end{align}
where $\beta = \frac{1}{i \Phi_F} - \epsilon$ for the wavefunction in the far future, $\beta = -\frac{1}{i \Phi_P} - \epsilon$ for the wavefunction in the far past, and $\{f(u),u\} = \frac{f'''(u)}{f'(u)} - \frac{3}{2} \left( \frac{f''(u)}{f'(u)}\right)^2$ is the Schwarzian derivative of $f(u)$ with respect to $u$.  Let us unpack the various ingredients in~\eqref{E:diskpathint1}.

First, the factor of $e^{S_0}$ in~\eqref{E:diskpathint1} comes from the Euler term in the action, $\chi = \frac{1}{4\pi} \int d^2 x \, \sqrt{-g}\,R + \frac{1}{2\pi} \int dx \, \sqrt{h}\,K$.  Evaluating this on the geometry~\eqref{E:HHfuture} which prepares and then evolves the no-boundary state (or its $(-,-)$ signature Euclidean AdS$_2$ continuation), we find $\chi = - i$.  As such, $e^{i S_0 \chi} = e^{S_0}$.  More generally, the Euler term in the action is related to the topological Euler characteristic of the underlying manifold by $\chi = -i \chi_T$, and so $e^{i S_0 \chi} = e^{S_0 \chi_T}$.

The field $f(u)$ in~\eqref{E:diskpathint1} is a $\text{Diff}(\mathbb{S}^1)$ field satisfying $f(u + 2\pi) = f(u) + 2\pi$ which characterizes reparameterizations of the asymptotic boundaries of our nearly-dS$_2$ spacetime~\cite{Maldacena:2019cbz, Cotler:2019nbi}.  It is weighted by the Schwarzian action familiar in JT gravity.  Thanks to the $\text{PSL}(2;\mathbb{R})$ isometry of the geometry~\eqref{E:HHfuture} we identify $f(u)$ modulo PSL$(2;\mathbb{R})$ transformations by $\tan\!\left( \frac{f(u)}{2}\right)\sim \frac{a \tan\left( \frac{f(u)}{2}\right)+b}{c\tan\left( \frac{f(u)}{2}\right)+d}$ where $ad-bc=1$. That is, $f(u)\in\faktor{\text{Diff}(\mathbb{S}^1)}{\text{PSL}(2;\mathbb{R})}$. The action and integration measure over $f(u)$ respect this identification.

Finally, the integration measure $\text{Pf}(-\Omega)$ in~\eqref{E:diskpathint1} is the Pfaffian of minus the symplectic form, which in this instance is the Kirillov-Kostant symplectic form on $\faktor{\text{Diff}(\mathbb{S}^1)}{\text{PSL}(2;\mathbb{R})}$~\cite{Witten:1987ty, Alekseev:1988ce}
\begin{align}
\label{E:Omegasymplectic1}
\Omega = \frac{1}{(2\pi)^2}\int_0^{2\pi} \! du \,\left[\frac{d f'(u) \wedge df''(u)}{f'^2(u)} - d f(u) \wedge df'(u)\right]\,,
\end{align}
where $df(u)$ is a formal one-form on the space of variations of $f(u)$.\footnote{We have chosen a particular normalization for the symplectic form here. We will comment more on this shortly.}

For positive $\beta$ the integral~\eqref{E:diskpathint1} has a saddle-point approximation and moreover is one-loop exact thanks to localization~\cite{Stanford:2017thb}.  We claim a similar result for slightly negative $\beta$. Let us take $\beta$ to have a very small negative real part. Modulo the PSL$(2;\mathbb{R})$ redundancy there is a unique extremum of the action given by $f(u)=u$. We expand in fluctuations around it, $f(u) = u + \varepsilon(u)$ and decompose $\varepsilon(u)$ into Fourier modes as $\varepsilon(u) = \sum_n e^{ -i n u}(\varepsilon_n^{(R)} + i \,\varepsilon_n^{(I)})$, with $\varepsilon_n^{(R)} = \varepsilon_{-n}^{(R)}$ and $\varepsilon_n^{(I)} = - \varepsilon_{-n}^{(I)}$. The PSL$(2;\mathbb{R})$ redundancy allows us to fix $\varepsilon_{n=-1,0,1}$ to vanish so that the sum only runs over $|n|\geq 2$. In terms of these fluctuations the symplectic form reads
\begin{align}
\Omega = \frac{2}{\pi} \sum_{n \geq 2}n (n^2 - 1)\, d\varepsilon_n^{(R)} \wedge d\varepsilon_{n}^{(I)} + O(\varepsilon^3)
\end{align}
so that
\begin{align}
	\text{Pf}(-\Omega) = \prod_{n\geq 2}\left(  (-1)\frac{2}{\pi}n(n^2-1) d\varepsilon_n^{(R)}d\varepsilon_n^{(I)} + O(\varepsilon^3)\right)\,.
\end{align}
Then the 1-loop approximation to~\eqref{E:diskpathint1} is 
\begin{align}
\label{E:diskpathint2}
	\widetilde{Z}_{\rm disk,1-loop}(\beta) = e^{S_0 + \frac{1}{\beta}} \prod_{n \geq 2} (-1)\frac{2}{\pi}(n^3-n)\int d\varepsilon_n^{(R)}\,d\varepsilon_n^{(I)}\,e^{-\frac{2}{\beta}n^2(n^2-1) \left((\varepsilon_n^{(R)})^2 + (\varepsilon_n^{(I)})^2\right)} \,.
\end{align}
The effect of taking the measure to be $\text{Pf}(-\Omega)$ rather than $\text{Pf}(\Omega)$ is to simply introduce a factor of $-1$ for each pair of modes. Nearly-dS$_2$ boundary conditions give $\beta = \frac{1}{i \Phi_F} - \epsilon$ for the future Hartle-Hawking wavefunction and $\beta = -\frac{1}{i \Phi_P} - \epsilon$ for its past counterpart, and so in each case we have $\text{Re}(\beta) < 0$. Then~\eqref{E:diskpathint2} is a product of wrong sign Gaussian integrals; to correct this, we rotate the integration contour of the $\varepsilon$'s, which has the effect of canceling the $-1$'s coming from the measure.\footnote{In Subsection~\ref{S:BF} we will see that the integration measure $\text{Pf}(-\Omega)$ is inherited from a negative-definite inner product of fluctuations on the reduced phase space, parameterized here by $\varepsilon$. Rotating the integration contour renders the inner product of fluctuations positive-definite, consistent with the ``rule of thumb'' of~\cite{Liu:2023jvm}. Indeed, all of the Schwarzian path integrals considered in this work are consistent with that prescription.} Then we obtain
\begin{align}
\label{E:preHH1}
	\widetilde{Z}_{\rm disk,1-loop}(\beta) = e^{S_0 + \frac{1}{\beta}} \prod_{n \geq 2}\frac{(-\beta)}{n}= \frac{1}{\sqrt{2\pi} (-\beta)^{3/2}}\,e^{S_0 + \frac{1}{\beta}}\,.
\end{align}
The second equation follows from a Zeta-regularization of the infinite product.  A direct evaluation of the two-loop contribution as in~\cite{Stanford:2017thb} shows that it vanishes. More generally, using the logic behind the localization argument of~\cite{Stanford:2017thb} in which one represents the Pfaffian with Grassmann-odd fields, notes that the action is closed with respect to a Grassmann-odd symmetry $Q$, and adds a $Q$-exact term to the action with a large coefficient, we find that $\widetilde{Z}_{\rm disk}$ receives no corrections to any order in perturbation theory in $\beta$. Thus
\begin{align}
\label{E:dSdisk}
	\widetilde{Z}_{\text{disk}}(\beta) = \frac{1}{\sqrt{2\pi} (-\beta)^{3/2}}\,e^{S_0 + \frac{1}{\beta}}\,.
\end{align}

For reference, the disk amplitude of Euclidean AdS JT gravity is, with these conventions,
\beq
\label{E:AdSdisk}
	Z_{\rm disk}(\beta) = \frac{1}{\sqrt{2\pi}\,\beta^{3/2}}\,e^{S_0+\frac{1}{\beta}}\,,
\eeq
with $\beta$ positive. The difference between~\eqref{E:dSdisk} and~\eqref{E:AdSdisk} is simply a replacement $\sqrt{\beta} \to \sqrt{-\beta}$ in the 1-loop prefactor, which amounts to an overall phase in the amplitude. Taking
$\beta = \frac{1}{i\Phi}- \epsilon$ corresponding to the future Hartle-Hawking wavefunction~\eqref{E:preHH1} gives us
\begin{align}
	\langle \Phi | \text{HH} \rangle_{g=0} =\frac{(-i \Phi)^{3/2}}{\sqrt{2\pi}}\, e^{S_0 + i \Phi}\,,
\end{align}
where $\Phi$ has a small positive imaginary part. The above agrees with previous results~\cite{Maldacena:2019cbz, Cotler:2019nbi} up to a global phase which was not fixed by prior work.  In those previous calculations, the overall phase depended on the definition of $\langle \Phi|$ and so was not emphasized.  In our context, the phase is important since we are choosing a convention for $\langle \Phi|$ once and for all and then computing higher-genus amplitudes for which overall phases are contentful.  In summary, we have found the identification
\begin{align}
\widetilde{Z}_{\text{disk}}\!\left(\frac{1}{i \Phi_F} - \epsilon\right) = Z_{\text{disk}}^{\text{dS}}(\Phi_F)
\end{align}
for the future wavefunction of the no-boundary state and similarly for the past wavefunction.

Now let us provide a similar analysis for the Schwarzian path integral describing a trumpet, which has the topology of an annulus. The relevant $(-,-)$ signature metric is
\beq
	ds^2 = - (d\rho^2 + b^2 \cosh^2(\rho) du^2)
\eeq
for $\rho \geq 0$. This geometry has a bottleneck at $\rho=0$ with length $2\pi b$.  The trumpet path integral reads
\begin{align}
\label{E:trumpetpathint1}
\widetilde{Z}_T(\beta,b) = \int \text{Pf}(-\Omega)\,e^{\frac{1}{\pi \beta} \int_0^{2\pi} du\, \left(\{f(u),u\} - \frac{b^2}{2} f'(u)^2\right)}
\end{align}
where as before $\beta = \frac{1}{i \Phi_F} - \epsilon$ for a future-directed trumpet and $\beta = - \frac{1}{i \Phi_P} - \epsilon$ for a past-directed trumpet. There is no $S_0$-dependence in the trumpet since $\chi = 0$ for the annulus.  Moreover $f(u)$ is a $\text{Diff}(\mathbb{S}^1)$ field satisfying $f(u + 2\pi ) = f(u) + 2\pi $. We also identify $f(u)\sim f(u) + \text{constant}$ on account of the $U(1)$ isometry of the geometry, so that $f(u)\in \faktor{\text{Diff}(\mathbb{S}^1)}{U(1)}$.

Performing a similar analysis as before, there is a unique saddle (modulo the $U(1)$ redundancy) $f(u)=u$, and expanding in small fluctuations around it as $f(u) = u + \varepsilon(u)$ we have
\begin{align}
	\Omega = \frac{2}{\pi} \sum_{n \geq 1} n\left(n^2 + b^2 \right) d\varepsilon_n^{(R)} \wedge d\varepsilon_n^{(I)} + O(\varepsilon^3)
\end{align}
so that
\begin{align}
	\text{Pf}(-\Omega) = \prod_{n \geq 1} \left( (-1)\frac{2}{\pi} n\left(n^2 + b^2 \right) d\varepsilon_n^{(R)}\,d\varepsilon_n^{(I)} + O(\varepsilon^3)\right)\,.
\end{align}
Then we can compute the 1-loop approximation to~\eqref{E:trumpetpathint1} (which is in fact exact by a similar argument as above)
\begin{align}
\label{E:trumpetpathint2}
	\widetilde{Z}_T(\beta,b) = e^{- \frac{b^2}{\beta}} \prod_{n \geq 1} (-1)\frac{2}{\pi}n(n^2 + b^2 )\int d\varepsilon_n^{(R)}\,d\varepsilon_n^{(I)}\,e^{-\frac{2}{\beta}n^2(n^2+b^2) \left((\varepsilon_n^{(R)})^2 + (\varepsilon_n^{(I)})^2\right) }\,.
\end{align}
Since $\text{Re}(\beta) < 0$, we again rotate the Schwarzian modes by $90^\circ$ and perform the integral over fluctuations, giving us
\begin{align}
\label{E:ZTtilde1}
\widetilde{Z}_T(\beta,b) = e^{- \frac{b^2}{\beta}} \prod_{n \geq 1} \frac{(-\beta)}{n} = \frac{1}{\sqrt{2\pi}(-\beta)^{1/2}}\,e^{- \frac{b^2}{\beta}}\,,
\end{align}
where we have Zeta-regularized the infinite product. For comparison, the standard Euclidean AdS trumpet with these conventions is $Z_T(\beta,b) = \frac{1}{\sqrt{2\pi}\, \beta^{1/2}}e^{-\frac{b^2}{\beta}}$, so as with the disk the dS trumpet is related to the AdS one by a simple replacement $\sqrt{\beta} \to \sqrt{-\beta}$ in the 1-loop prefactor.

Alternatively, we can consider a future Lorentzian trumpet geometry
\beq
\label{E:LorentzianTrumpet}
	ds^2 = -dt^2 + \alpha^2 \cosh^2(t) du^2
\eeq
for $t\geq 0$, with future dS JT boundary conditions. Here $\alpha$ parameterizes the bottleneck of the geometry at $t=0$, which has a length $2\pi \alpha$. The dS JT path integral for such a space is
\beq
	Z_T^{\rm dS}(\Phi_F,\alpha) = \int \text{Pf}(-\Omega) e^{\frac{i \Phi_F}{\pi}\int_0^{2\pi} du\left( \{f(u),u)\} + \frac{\alpha^2}{2} f'(u)^2\right)}\,,
\eeq
where $\Phi_F$ has a slightly positive imaginary part. The computation of this path integral is nearly identical to that above, with the result
\beq
\label{E:dStrumpet}
	Z_T^{\rm dS}(\Phi_F,\alpha) =\frac{(-i\Phi_F)^{1/2} }{\sqrt{2\pi}}e^{i \Phi_F \alpha^2} = \widetilde{Z}_T\left( \beta = \frac{1}{i\Phi_F} - \epsilon,b=i \alpha\right)\,,
\eeq
i.e. it is merely the continuation of the $(-,-)$ trumpet both in $\beta$ and in $b$ with $\alpha^2 = - b^2$. We can similarly consider a past Lorentzian trumpet, given by the $t\leq 0$ part of~\eqref{E:LorentzianTrumpet}, with past dS JT boundary conditions in terms of some $\Phi_P$ with small negative imaginary part. It is given by the complex conjugate of the future trumpet, namely
\beq
\label{E:dStrumpet2}
	Z_T^{\rm dS*}(\Phi_P,\alpha) = \frac{(i \Phi_P)^{1/2}}{\sqrt{2\pi}}e^{-i \Phi_P \alpha^2} = \widetilde{Z}_T \left( \beta =-\frac{1}{i \Phi_P} - \epsilon,b=i \alpha\right)\,.
\eeq
Notice that because of our $i\epsilon$ prescription wherein $\Phi_F$ has a small positive imaginary part and $\Phi_P$ has a small negative imaginary part, the trumpets~\eqref{E:dStrumpet} and~\eqref{E:dStrumpet2} are damped at very large $\alpha^2$.

In~\cite{Cotler:2019nbi} similar de Sitter trumpets were considered. The trumpets in~\cite{Cotler:2019nbi} are the same as those above up to a phase coming from the JT path integral measure, and a careful treatment of $i\epsilon$'s.

The de Sitter trumpet has the physical interpretation~\cite{Cotler:2023eza} as a transition amplitude between a ``bottleneck'' state of fixed size $|\alpha\rangle$ of the bulk Hilbert space and a state $|\Phi\rangle$ in the space of states prepared at infinity. In particular, we identify to leading order in the topological expansion
\beq
	Z_T^{\rm dS}(\Phi_F,\alpha) = \langle \Phi_F|\widehat{V}|\alpha\rangle\,, \qquad Z_T^{\rm dS*}(\Phi_P,\alpha) = \langle \alpha |\widehat{V}^{\dagger}| \Phi_P\rangle\,,
\eeq
where $\widehat{V}$ is the evolution operator from the bulk to asymptotic future infinity, and its conjugate $\widehat{V}^\dagger$ is the evolution operator from past infinity to the bulk.

\subsection{Amplitudes}

We are now in a position to compute the dS JT path integral $\widetilde{Z}_{g,n}(\beta_1,..,\beta_n)$ on a general surface of genus $g$ with $n$ asymptotic circles. Let us begin with the cylinder $\widetilde{Z}_{0,2}$, and then progress to the general case.

Before doing so, let us briefly comment on our conventions. In the notation of Saad, Shenker, and Stanford~\cite{Saad:2019lba}, we normalize the symplectic form with $\alpha_{\rm them} = \frac{1}{2\pi^2}$; choose dilaton boundary conditions so that the Schwarzian action associated with an asymptotic boundary has $\gamma_{\rm them} = \frac{1}{2\pi^2}$; pick a different normalization for the bottleneck parameter, $b_{\rm them} = 2\pi b_{\rm us}$; and normalize our twist moduli that parameterize the gluing of trumpets to the intermediate surface as $\tau_{\rm them} = \frac{\tau_{\rm us}}{2\pi}$.

\subsubsection{Cylinder and inner product}
\label{subsub:cyl}

We warm up with the cylinder or ``double trumpet,'' which we can compute as a sum over $(-,-)$ hyperbolic metrics or as a sum over $(-,+)$ metrics, with identical results. We begin with the sum over $(-,-)$ metrics, which can be represented as
\beq
	ds^2 = -\left( d\rho^2 + b^2 \cosh^2(\rho)(dx + \tau \delta(\rho) d\rho)^2\right)\,.
\eeq
Here $b>0$ and $\tau\sim \tau+2\pi$ are the moduli of the hyperbolic cylinder. This sum may be performed by regarding the cylinder as two trumpets, one with $\rho>0$ and the other with $\rho<0$, soldered together at the bottleneck $\rho=0$ up to a twist $\tau$, together with an appropriate integral over the moduli. The symplectic form restricted to variations of $b$ and $\tau$ is $\Omega = \frac{db^2 \wedge d\tau}{2\pi}$ so that the integration measure over $(b,\tau)$ inherited from $\text{Pf}(-\Omega)$ is $-\frac{db^2 d\tau}{2\pi}$. The cylinder amplitude is thus
\beq
	\widetilde{Z}_{0,2}(\beta_1,\beta_2) =\int_0^{2\pi} \frac{d\tau}{2\pi} \int_0^{\infty} (-db^2) \widetilde{Z}_T(\beta_1,b)\widetilde{Z}_2(\beta_2,b) = -\int_0^{\infty} db^2 \,\frac{1}{2\pi \sqrt{-\beta_1}\sqrt{-\beta_2}} e^{-b^2 \left( \frac{1}{\beta_1}+\frac{1}{\beta_2}\right)}\,.
\eeq
Because $\beta_1$ and $\beta_2$ have negative real parts, we rotate the $b^2$ contour of integration as $b^2 = -\alpha^2$ so that\footnote{We can think of the ensuing integral over $\alpha$ as an integral over Lorentzian bottlenecks.}
\beq
\label{E:doubleTrumpet}
	\widetilde{Z}_{0,2}(\beta_1,\beta_2) = \int_0^{\infty} d\alpha^2 \frac{1}{2\pi\sqrt{-\beta_1}\sqrt{-\beta_2}}e^{\alpha^2\left( \frac{1}{\beta_1}+\frac{1}{\beta_2}\right)} = - \frac{1}{2\pi} \frac{\sqrt{-\beta_1}\sqrt{-\beta_2}}{\beta_1+\beta_2}\,.
\eeq

Suppose that we are interested in the global dS$_2$ amplitude, with a future circle characterized by $\Phi_F$ and a past circle characterized by $\Phi_P$. The relevant $R=2$ spacetimes are 
\beq
\label{E:globaldS}
	ds^2 = -dt^2 + \alpha^2\cosh^2(t) (du + \tau \delta(t)dt)^2\,,
\eeq
where $\alpha > 0$ and $\tau \sim \tau+2\pi$ label the moduli. This is a topological cylinder, and we can obtain this Lorentzian amplitude from a continuation of the cylinder amplitude under $\beta_1 = \frac{1}{i\Phi_F} - \epsilon$ and $\beta_2 = -\frac{1}{i \Phi_P} - \epsilon$. With this continuation $\widetilde{Z}_{0,2}(\beta_1,\beta_2)$ becomes
\beq
\label{E:globalAmplitude}
	\widetilde{Z}_{0,1,1}(\Phi_F;\Phi_P) = \frac{i}{2\pi} \frac{\sqrt{\Phi_F}\sqrt{ \Phi_P} }{\Phi_F -\Phi_P + i \epsilon}\,.
\eeq
Here $\widetilde{Z}_{g,p,q}$ refers to the dS JT amplitude for a genus $g$ surface with $p$ future asymptotic circles and $q$ past ones. 

We can also obtain the result~\eqref{E:globalAmplitude} by directly summing over Lorentzian $R=2$ metrics~\eqref{E:globaldS} where we consider a past dS$_2$ trumpet characterized by a bottleneck size $\alpha$ and glue it to a future dS$_2$ trumpet, with a moduli space measure $\frac{d\alpha^2 d\tau}{2\pi}$ inherited from $\text{Pf}(-\Omega)$ via $-\frac{db^2d\tau}{2\pi}$.  The resulting amplitude is
\beq
	\widetilde{Z}_{0,1,1}(\Phi_F;\Phi_P) = \int_0^{2\pi} \frac{d\tau}{2\pi} \int_0^{\infty} d\alpha^2 \,Z_T^{\rm dS}(\Phi_F,\alpha)Z_T^{\rm dS*}(\Phi_P,\alpha)\,.
\eeq

A similar Lorentzian signature computation using the same measure gives the inner product on asymptotic states~\cite{Cotler:2019dcj}. The inspiration for that computation is the fact that the zero-time limit of transition amplitudes limits to the inner product. For two asymptotic states, we sum over cylinders that connect a large circle in the asymptotic future with some $\Phi'$ (which prepares the ket) to another large circle just after it with some $\Phi$ (which prepares the bra). The JT path integral over such a cylinder is a moduli space integral over the same integrand as above, but where $\alpha^2$ is integrated over the whole real line. The reason is that, since these cylinders are in the far future part of the geometry~\eqref{E:globaldS}, they are non-singular for any $\alpha^2$ on the real line.\footnote{After a coordinate transformation the line element~\eqref{E:globaldS} is, in the far future $ds^2 \approx -dt^2 + \left(\frac{e^{2t}}{4}+ 2 \alpha^2\right)du^2$.} The inner product is then
\begin{align}
\label{E:innerproduct1}
\langle \Phi | \Phi'\rangle = \int_0^{2\pi}\frac{d\tau}{2\pi}\int_{-\infty}^\infty d\alpha^2 \, Z_T^{\text{dS}}(\Phi,\,\alpha) \, Z_{T}^{\text{dS}\,*}(\Phi', \alpha) = \sqrt{\Phi} \sqrt{\Phi'}\, \delta(\Phi - \Phi')\,,
\end{align}
which is positive-definite on account of the symplectic measure $\text{Pf}(-\Omega)$. If we had instead used $\text{Pf}(\Omega)$ we would have ended up with minus the above result. The positive-definiteness of~\eqref{E:innerproduct1} is the main reason why we choose the measure $\text{Pf}(-\Omega)$.

Because the inner product~\eqref{E:innerproduct1} is non-trivial, the JT path integral produces an unnormalized version of the $S$-matrix. Rescaling the external states as \footnote{This is slightly different than in our previous work~\cite{Cotler:2019dcj} where we rescaled by $1/\sqrt{\Phi}$.} $|\Phi\rangle \to \frac{|\Phi\rangle}{\sqrt{i\Phi}}$, we arrive at the properly normalized $S$-matrix. The properly normalized global dS$_2$ amplitude, which describes the amplitude for a large past universe to evolve into a large future one (where we project out the possibility that the past circle evolves into the no-boundary state, and that the future circle arises from another no-boundary state), becomes
\beq
	\langle \Phi | \,\widehat{\mathcal{U}}\,| \Phi'\rangle \simeq \frac{i}{2\pi}\frac{1}{\Phi - \Phi' + i \epsilon} + O(e^{-2S_0})\,,
\eeq
where $\widehat{\mathcal{U}}$ is the infinite-time evolution operator. Treating $\widehat{\Phi}$ as a ``position'' and introducing a canonically conjugate ``momentum'' $\widehat{p}$, the above amplitude is, in the momentum basis,
\beq
	\langle p | \,\widehat{\mathcal{U}}\,|p'\rangle \approx \Theta(p) \delta(p-p')\,.
\eeq 
This equation implies that $\widehat{\mathcal{U}}$ is approximately a projector as opposed to a unitary.  A more detailed analysis of bulk versus asymptotic Hilbert spaces~\cite{Cotler:2023eza} establishes that $\widehat{\mathcal{U}} = \widehat{V}\widehat{V}^{\dagger}$ where $\widehat{V}$ is the semi-infinite evolution from the bulk to future asymptotic infinity, and $\widehat{V}^{\dagger}$ the evolution operator from past asymptotic infinity to the bulk, with $\widehat{V}$ an isometry and $\widehat{V}^{\dagger}$ a co-isometry.

\subsubsection{General surfaces}

The cylinder amplitude has the advantage that it can be computed from a suitable sum over Lorentzian $R=2$ metrics or over hyperbolic metrics in $(-,-)$ signature, each giving the same result. A general surface $\Sigma$ of genus $g$ with $n$ boundaries however does not admit a smooth $R=2$ Lorentzian metric, and so instead we can perform a sum over $(-,-)$ metrics. Fortunately, we have all of the ingredients necessary to do that sum.

By a cut-and-paste procedure we build up the moduli space of hyperbolic metrics on $\Sigma$ by dividing it into $n$ trumpets glued to an intermediate surface $\Sigma_0$ of genus $g$ across bottlenecks of lengths $2\pi b_i$. To the $i$th trumpet we associate $\widetilde{Z}_T(\beta_i,b_i)$ and include a measure $-\frac{db^2_i d\tau_i}{2\pi}$ over the moduli $(b_i,\tau_i)$ that parameterize how the trumpet is glued to $\Sigma_0$. Using the symplectic measure $\text{Pf}(-\Omega)$, the intermediate surface has, in the conventions of~\cite{Saad:2019lba}, a symplectic volume $(-1)^{3g-3+n} V_{g,n}^{\alpha}(2\pi b_1,\hdots,2\pi b_n)$ with $V_{g,n}^{\alpha}$ the Weil-Petersson volume of the moduli space of hyperbolic metrics on a genus $g$ surface $\Sigma_0$ with $n$ geodesic boundaries of lengths $2\pi b_i$. The factor of $(-1)^{3g-3+n}$ arises because there are $3g-3+n$ pairs of moduli on $\Sigma_0$. The parameter $\alpha$ appearing in~\cite{Saad:2019lba}, not to be confused with the modulus of global dS$_2$ in~\eqref{E:globaldS}, controls the normalization of the symplectic form, and in our conventions is $\frac{1}{2\pi^2}$. In what follows we drop this $\alpha$ label out of convenience. 

The amplitude is then
\begin{align}
\begin{split}
\label{E:masterAmplitude0}
	\widetilde{Z}_{g,n}(\beta_1,...,\beta_n) = (-1)^{3g-3} \int_0^{\infty} db_1^2 \cdots db_n^2 \,V_{g,n}(2\pi b_1,...,2\pi b_n) \widetilde{Z}_T(\beta_1,b_1) \cdots \widetilde{Z}_T(\beta_n,b_n)\,.
\end{split}
\end{align}
Because the $\beta_i$ have negative real parts, the integrals do not converge. Away from infinity the integrand is an analytic function of the $b_i^2$ and so we deal with this by rotating the $b_i^2$ contours via $b_i^2 = -\alpha_i^2$ so that\footnote{The Weil-Petersson volumes are even polynomials in the $b_i$ and so this continuation is unambiguous.}
\beq
\label{E:masterAmplitude}
	\widetilde{Z}_{g,n}(\beta_1,...,\beta_n) = (-1)^{3g-3+n} \int_0^{\infty} d\alpha_1^2 \cdots d\alpha_n^2 \, V_{g,n}(2\pi i \alpha_1,...,2\pi i \alpha_n) \widetilde{Z}_T(\beta_1,i \alpha_1) \cdots \widetilde{Z}_T(\beta_n,i \alpha_n)\,.
\eeq
Analytically continuing the $\beta$'s to represent $p$ future circles labeled by $\{\Phi_a\}$ and $q=n-p$ past circles with $\{\Phi_m'\}$ we arrive at an expression for the Lorentzian transition amplitudes
\begin{align}
\label{E:masterAmplitude2}
	\widetilde{Z}_{g,p,q}(\Phi_1,...,\Phi_p;\Phi'_1,...\Phi'_q) & =(-1)^{3g-3+n} \int_0^{\infty} d\alpha_1^2 \cdots d\alpha_n^2 \,V_{g,n}(2\pi i \alpha_1,...,2\pi i \alpha_n) 
	\\
	\nonumber
	&\qquad \qquad \quad \! \times Z_T^{\rm dS}(\Phi_1,\alpha_1) \cdots Z_T^{\rm dS} (\Phi_p,\alpha_p) Z_T^{\rm dS*}(\Phi_1',\alpha_{p+1}) \cdots Z_T^{\rm dS*} (\Phi_q',\alpha_n)\,.
\end{align}
This expression is quite nearly the result that was argued for in~\cite{Cotler:2019nbi} using a very different picture for the spacetime, with de Sitter trumpets glued to a Euclidean surface with cone points, along with a (ultimately incorrect) conjecture concerning the Weil-Petersson volume for surfaces with cone points~\cite{do2009weil,Turiaci:2020fjj,Eberhardt:2023rzz}. The differences, which we have found by a careful treatment of the measure and the $i\epsilon$ prescription, are the phases associated with the trumpets, and the factor of $(-1)^{3g-3+n}$. The integration over $\alpha_i$'s arises out of the need to define a convergent moduli space integral over hyperbolic metrics with negative $\beta$'s.

So far we have dealt with the disk, the cylinder, and general surfaces of genus $g$ and $n$ boundaries. An exception to this analysis is the sphere, which has no moduli and no boundaries, and thus no Schwarzian modes. The sphere amplitude is just a constant, which diverges owing to a triplet of dilaton zero modes with non-compact field range, i.e.
\beq
	\widetilde{Z}_{\mathbb{S}^2} = e^{2S_0}\times \infty\,.
\eeq
This divergence ought to be expected. Recall that the genus zero approximation to the Hartle-Hawking wavefunction of the no-boundary state is, after normalizing the asymptotic state $\langle \Phi|$,
\beq
	\Psi_{\rm HH}(\Phi) =\langle \Phi|\text{HH}\rangle \approx -\frac{i \Phi}{\sqrt{2\pi}} e^{S_0 + i \Phi}\,.
\eeq
Thinking of the Hartle-Hawking wavefunction as a matrix element of the semi-infinite evolution operator $\widehat{V}$ between the no-boundary state $|\varnothing\rangle$ and an asymptotic state,
\beq
	\Psi_{\rm HH}(\Phi) = \langle \Phi|\widehat{V}|\varnothing\rangle \,,
\eeq
the normalization of the Hartle-Hawking state is
\beq
	\langle \text{HH}|\text{HH}\rangle =\langle \varnothing|\widehat{V}^{\dagger}\widehat{V}|\varnothing\rangle\,.
\eeq
Since $\widehat{V}$ obeys $\widehat{V}^{\dagger}\widehat{V} \approx \mathds{1}$ to leading order in the genus expansion, we then have
\beq
	\langle \text{HH}|\text{HH}\rangle \approx \langle \varnothing|\varnothing\rangle \approx \widetilde{ Z}_{\mathbb{S}^2}\,.
\eeq
But
\beq
\langle \text{HH}|\text{HH}\rangle \approx   \int_{-\infty}^{\infty} d\Phi \,|\Psi_{\rm HH}(\Phi)|^2 \approx \frac{e^{2S_0}}{2\pi} \int_{-\infty}^{\infty} d\Phi \,\Phi^2 = e^{2S_0}\times \infty\,,
\eeq
which matches the divergent sphere partition function.  While the sphere partition function of JT gravity diverges, we note that it is finite for $(2,2p-1)$ minimal strings at finite $p$~\cite{Anninos:2021ene,Mahajan:2021nsd}.

\subsubsection{Relation to Euclidean AdS}
\label{S:AdStodS}

Since both Euclidean AdS JT gravity and dS JT gravity involve a sum over hyperbolic metrics, it is natural to expect a relation between dS amplitudes and their Euclidean AdS cousins. 

Consider the general expression~\eqref{E:masterAmplitude0} for the dS JT path integral on a surface $\Sigma_{g,n}$. The corresponding expression with the same conventions for the normalization of the symplectic form and the Schwarzian action is
\beq
\label{E:AdSamplitudes}
	Z_{g,n}(\beta_1,...,\beta_n) = \int_0^{\infty} db_1^2 \cdots db_n^2 \,V_{g,n}(2\pi b_1,...,2\pi b_n) Z_T(\beta_1,b_1) \cdots Z_T(\beta_n,b_n)\,,
\eeq
where $Z_T(\beta,b) = \frac{1}{\sqrt{2\pi \beta}}e^{-\frac{b^2}{\beta}} = \frac{\sqrt{-\beta}}{\sqrt{\beta}}\widetilde{Z}_T(\beta,b)$ is the Euclidean AdS trumpet. Using that the $V_{g,n}$'s are even polynomials and performing the moduli space integrals, we see that
\begin{align}
\begin{split}
\mathcal{Z}_{g,n}(\beta_1,...,\beta_n) &:=\sqrt{\beta_1} \cdots \sqrt{\beta_n} \,Z_{g,n}(\beta_1,...,\beta_n)\,,
\\
\widetilde{\mathcal{Z}}_{g,n}(\beta_1,...,\beta_n) & : = \sqrt{-\beta_1} \cdots \sqrt{-\beta_n} \,\widetilde{Z}_{g,n}(\beta_1,...,\beta_n)\,,
\end{split}
\end{align}
are polynomials in the $\beta$'s.  Since they are analytic functions apart from singularities at $\infty$, they can be analytically continued for all $\beta$. Comparing their analytic continuations we land on the simple relation
\beq
\label{E:continuation}
	\widetilde{\mathcal{Z}}_{g,n}(\beta_1,...,\beta_n) = (-1)^{3g-3} \mathcal{Z}_{g,n}(\beta_1,...\beta_n)\,.
\eeq
Comparing the dS disk amplitude~\eqref{E:dSdisk} with the AdS version~\eqref{E:AdSdisk} and comparing the dS cylinder amplitude~\eqref{E:doubleTrumpet} with the AdS version $Z_{0,2}(\beta_1,\beta_2) = \frac{1}{2\pi} \frac{\sqrt{\beta_1}\sqrt{\beta_2}}{\beta_1+\beta_2}$, we see that the relation~\eqref{E:continuation} is also obeyed for the disk and cylinder.

Eq.~\eqref{E:continuation} expresses that dS JT amplitudes are the suitable continuation of AdS JT amplitudes.  Let us give two examples. The first is the ``triple trumpet'' with $g=0$ and $n=3$. The Weil-Petersson volume in this case is $V_{0,3} = 1$ and so
\begin{align}
\begin{split}
	\widetilde{Z}_{0,3}(\beta_1,\beta_2,\beta_3) &= -\int_0^{\infty} db_1^2db_2^2db_3^2 \, \widetilde{Z}_T(\beta_1,b_1) \widetilde{Z}_T(\beta_2,b_2) \widetilde{Z}_T(\beta_3,b_3) 
	\\
	& = - \frac{\beta_1\beta_2\beta_3}{(2\pi)^{3/2}\sqrt{-\beta_1}\sqrt{-\beta_2}\sqrt{-\beta_3}}\,,
\end{split}
\end{align}
which produces $\widetilde{\mathcal{Z}}_{0,3}(\beta_1,\beta_2,\beta_3) = - \frac{\beta_1\beta_2\beta_3}{(2\pi)^{3/2}} = - \mathcal{Z}_{0,3}(\beta_1,\beta_2,\beta_3)$. The second example is when the surface has genus 1 with a single boundary. The relevant Weil-Petersson volume is (accounting for the fact that, in the conventions of~\cite{Saad:2019lba}, the normalization constant $\alpha$ is $\frac{1}{2\pi^2}$)
\beq
	V_{1,1}(2\pi b) = \frac{b^2+1}{24}\,,
\eeq
and so
\begin{align}
\begin{split}
	\widetilde{Z}_{1,1}(\beta) & = \int_0^{\infty} db^2\left( \frac{b^2+1}{24}\right) \widetilde{Z}_T(\beta,b)
	\\
	& = \frac{\beta(\beta+1)}{24\sqrt{2\pi}\sqrt{- \beta}}\,,
\end{split}
\end{align}
which produces $\widetilde{\mathcal{Z}}_{1,1}(\beta) = \frac{\beta(\beta+1)}{24\sqrt{2\pi}} = \mathcal{Z}_{1,1}(\beta)$.

Including the genus counting parameter $e^{-S_0}$, we arrive at the genus expansion of the connected $n$-boundary amplitude of dS JT gravity,
\begin{align}
\label{E:dSgenus}
\begin{split}
\centering
\includegraphics[scale=.3]{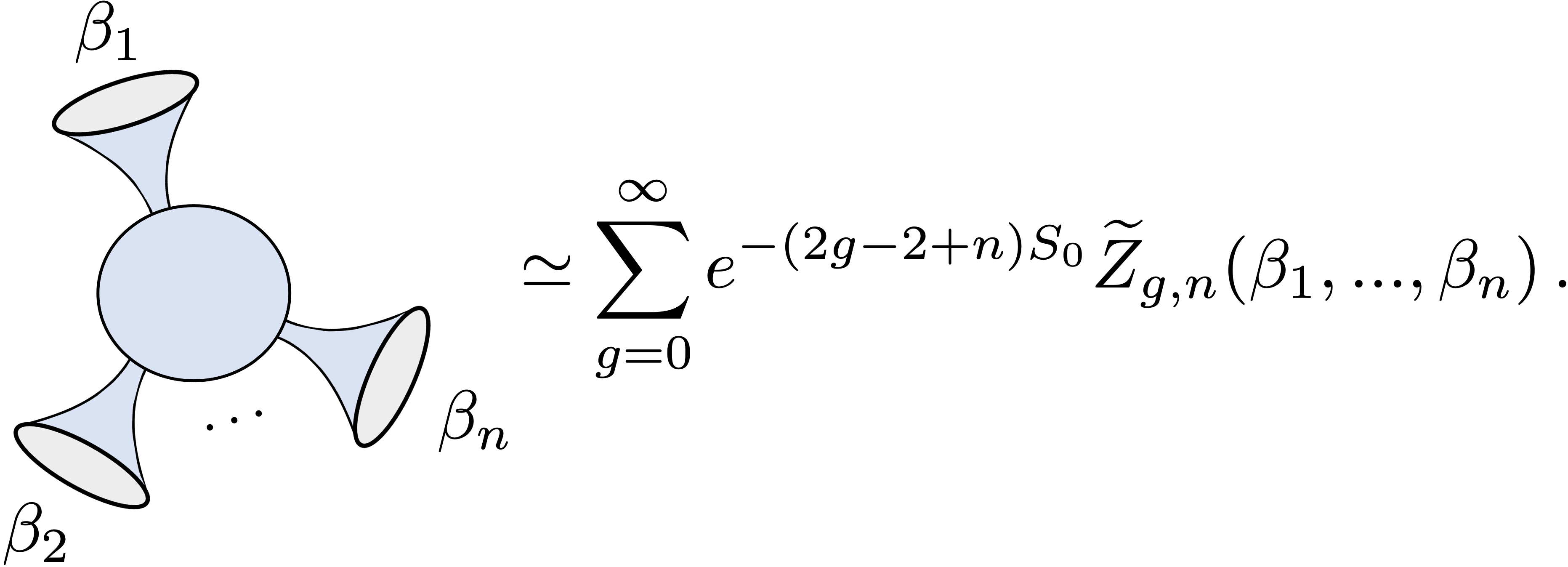}
\end{split}
\end{align}
The same expression with the replacement $\widetilde{Z}_{g,n}\to Z_{g,n}$ describes the genus expansion of the $n$-boundary amplitude of AdS JT gravity. We use a $\simeq$ rather than an $=$ because this is merely a series in $e^{-S_0}$, which for AdS JT gravity is asymptotic and moreover does not account for effects which are not analytic functions of $e^{-S_0}$. Because $\widetilde{Z}_{g,n}$ has an alternating sign $(-1)^{3g-3}$ we immediately learn that dS JT has an alternating sign genus expansion, in contrast with the same sign expansion of its AdS cousin.

In fact the genus expansion of dS JT gravity is a suitable analytic continuation of the AdS JT genus expansion as we now show. Take the genus-$g$ contribution to the $n$-boundary problem in AdS JT gravity, $e^{-(2g-2+n)S_0}Z_{g,n}(\beta_1,...,\beta_n)$. Analytically continuing the $\beta$'s to the complex plane and passing to $\text{Re}(\beta)<0$ by going counterclockwise around the origin, we have $\sqrt{\beta_j} \to i \sqrt{-\beta_j}$. We combine this continuation with an additive additive renormalization of $S_0\to \widetilde{S}_0 = S_0 + i \frac{3\pi}{2}$, under which $e^{-(2g-2+n)S_0} \to (-1)^{3g-3} i^n e^{-(2g-2+n)S_0}$.\footnote{Equivalently, we could continue to $\text{Re}(\beta_J)<0$ by passing clockwise around the origin and send $S_0 \to S_0 - i\frac{3\pi}{2}$.} Then thanks to~\eqref{E:continuation} we have for $\text{Re}(\beta_i)<0$
\beq
\label{E:AdStodS}
	e^{-(2g-2+n)\widetilde{S}_0} Z_{g,n}(\beta_1,...,\beta_n)= e^{-(2g-2+n)S_0}\widetilde{Z}_{g,n}(\beta_1,...,\beta_n)\,,
\eeq
where the right-hand side provides the expansion coefficients of dS JT gravity.

To summarize and highlight our results, the dS genus expansion~\eqref{E:dSgenus} is obtained from the AdS genus expansion under 
\begin{enumerate}
	\item An additive imaginary renormalization of $S_0 \to S_0 + i\frac{3\pi}{2}$. The effective string coupling $g_s = e^{-S_0} \to i e^{-S_0}$ becomes pure imaginary, leading to an alternating sign genus expansion.
	\item A continuation from $\text{Re}(\beta)>0$ to $\text{Re}(\beta)<0$ through a counterclockwise trajectory around $\beta = 0$.
\end{enumerate}
In Subsection~\ref{S:dSAiry} we present evidence that these seemingly distinct operations are in fact closely connected.

\subsubsection{Extracting quantum mechanics}
\label{S:dSQM}

With the genus expansion~\eqref{E:masterAmplitude2} for dS JT amplitudes in hand we would like to extract some physics from it. We have already discussed the interpretation of the disk and cylinder amplitudes, the former computing the genus-0 approximation to the Hartle-Hawking wavefunction of the no-boundary state, and the latter the amplitude (where we project out evolution into and out of the no-boundary state) for an asymptotic circle in the far past to evolve to an asymptotic circle in the far future. What about the rest of the genus expansion?

To begin we must deduce the genus expansion of the normalized $S$-matrix. Since normalized asymptotic states are obtained by rescaling $|\Phi\rangle \to \frac{|\Phi\rangle}{\sqrt{i\Phi}}$ we have 
\beq
	\langle \Phi_1,...,\Phi_p|\,\widehat{\mathcal{U}}\,|\Phi_1',...,\Phi_q'\rangle_{\rm conn} \simeq \sum_{g=0}^{\infty} e^{-(2g-2+n)S_0} S_{g,p,q}(\Phi_1,...,\Phi_p;\Phi_1',...,\Phi_q')\,,
\eeq
where the expansion coefficients $S_{g,p,q}$ for a surface other than the disk and cylinder are given by
\begin{align}
	S_{g,p,q}(\Phi_1,...,\Phi_p;\Phi_1',...,\Phi_q') = (-1)^{3g-3+n} &\int_0^{\infty} d\alpha_1^2 \cdots d\alpha_n^2 \,V_{g,n}(2\pi i \alpha_1,...,2\pi i \alpha_n)
	\\
	\nonumber
	& \times S_T(\Phi_1,\alpha_1) \cdots S_T(\Phi_p,\alpha_p) S_T^*(\Phi_1',\alpha_{p+1}) \cdots S_T^*(\Phi_q',\alpha_n)\,,
\end{align}
with
\beq
	S_T(\Phi,\alpha) = \frac{1}{\sqrt{2\pi}}\,e^{i \Phi \alpha^2}\,.
\eeq
The normalized future disk and cylinder are
\begin{align}
\begin{split}
	S_{0,1,0}(\Phi) & = - \frac{i \Phi}{\sqrt{2\pi}}\,e^{i \Phi}\,,
	\\
	S_{0,1,1}(\Phi;\Phi') & = \frac{i}{2\pi} \frac{1}{\Phi-\Phi'+i \epsilon}\,.
\end{split}
\end{align}
The past disk is $S_{0,0,1}(\Phi) = S_{0,1,0}(\Phi)^*$ and the cylinder with two future or two past circles is given by the appropriate continuations of $\widetilde{Z}_{0,2}$ in~\eqref{E:doubleTrumpet}. 

In our discussion of the cylinder amplitude we found it convenient to pass from the $\widehat{\Phi}$ basis to the $\widehat{p}$ basis with $\widehat{p}$ the ``momentum'' conjugate to the ``position'' $\widehat{\Phi}$. Then $S_T(\Phi,\sqrt{p}) = \langle \Phi|p\rangle$. In the $p$ basis we have
\beq
	\langle p_1,...,p_p|\,\widehat{\mathcal{U}}\,| p'_1,...,p_q'\rangle_{\rm conn} \simeq \sum_{g=0}^{\infty} e^{-(2g-2+n)S_0} \widetilde{S}_{g,p,q}(p_1,...,p_p;p_1',...,p_q')\,,
\eeq
where for a general surface
\beq
	\widetilde{S}_{g,p,q}(p_1,...,p_p;p_1',...,p_q') = (-1)^{3g-3+n} V_{g,n}(2\pi i \sqrt{p_1},...,2\pi i \sqrt{p_q'}) \Theta(p_1) \cdots\Theta( p_p)\Theta(p_1')\cdots\Theta(p_q')\,,
\eeq
with $n=p+q$, and for the disk and cylinder
\beq
	\widetilde{S}_{0,1,0}(p) = \delta'(p-1)\,, \qquad \widetilde{S}_{0,1,1}(p,p') = \Theta(p)\delta(p-p')\,.
\eeq
The genus corrections in this basis are (up to a phase) simply the analytically continued Weil-Petersson volumes.

We are now in a position to extract interesting physics. With no boundaries in the past, we are dealing with the Hartle-Hawking wavefunction of the no-boundary state at future infinity, including all of the genus corrections. More generally we have transition amplitudes from an initial state characterized by $q$ asymptotic circles in the infinite past to a future state characterized by $p$ asymptotic circles in the infinite future.

In~\cite{Cotler:2023eza} we identified the bulk Hilbert space of dS JT gravity associated with a single closed universe. A suitable basis of states is given by the span of $|P\rangle$ with $P>0$ and where
\beq
\label{E:theobject1}
	\langle \Phi|\widehat{V}|P\rangle_{g=0} = \frac{1}{\sqrt{2\pi}}\,e^{i\Phi P}\,,
\eeq
with $\widehat{V}$ being the semi-infinite evolution operator from the bulk to the infinite future. The object~\eqref{E:theobject1} is just the de Sitter trumpet $Z_T^{\rm dS}$ in~\eqref{E:dStrumpet} upon normalizing the final state $\langle \Phi|$. At genus 0 there is also an inner product of bulk states $\langle P|P'\rangle_{g=0} = \delta(P-P')$.

\begin{figure}[t!]
\begin{center}
\includegraphics[scale=.55]{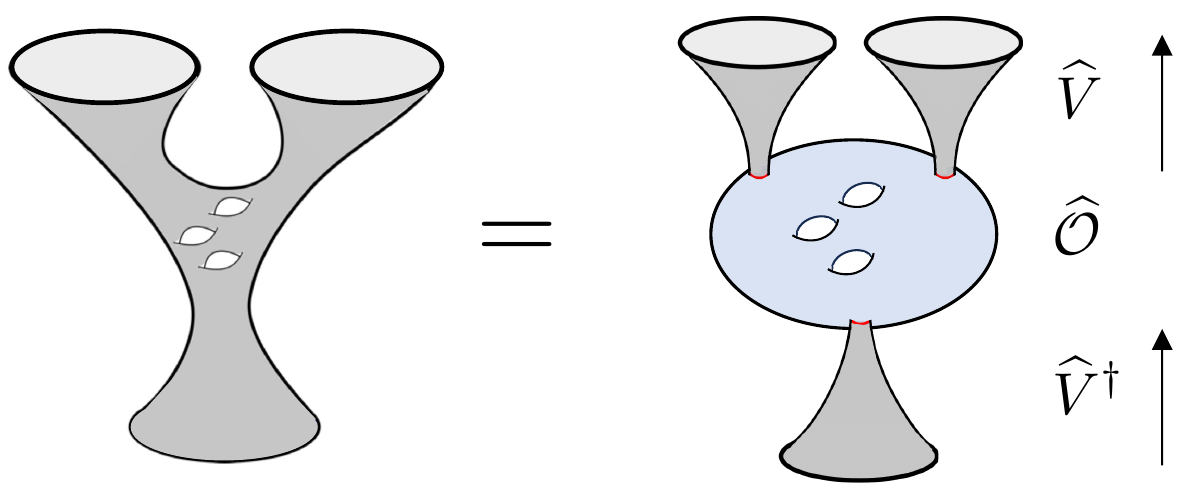}
\caption{\label{F:dSquantum} A conceptual depiction of the amplitude $S_{g,p,q}$. On the left-hand side, we represent the amplitude as a sum over $(-,-)$ metrics on a genus $g$ surface with $n=p+q$ boundaries. On the right-hand side we interpret that amplitude as an initial state of $q$ large circles in the far past evolved to the bulk using $\widehat{V}^{\dagger}$. Then time evolution acts with a bulk operator $\widehat{\mathcal{O}}$ that maps the $q$ incoming universes to $p$ outgoing universes, which are then evolved forward using $\widehat{V}$. }
\end{center}
\end{figure}

Taking these results together, we propose the following physical interpretation of the $p\to q$ $S$-matrix with $p,q>0$. Given the connected part of that $S$-matrix, we regard the evolution as split up into three steps:
\begin{enumerate}
	\item First we evolve from the asymptotic past to a bulk time using past de Sitter trumpets. An initial state $|p'_1,...,p_q'\rangle$ evolves to $\widehat{V}^{\dagger}|p_1',...,p_q'\rangle = \Theta(p_1')...\Theta(p_q')|P_1' = p_1',...,P_q'=p_q'\rangle$.
	\item There is a bulk overlap $\widehat{\mathcal{O}}$ from the bulk Hilbert space to itself that maps incoming states $|P_1',...,P_q'\rangle$ with $q$ closed universes to outgoing states $|P_1,...,P_p\rangle$ with $p$ closed universes. This operator has the genus expansion
	\beq
		\langle P_1,...,P_p|\,\widehat{\mathcal{O}}\,|P_1',...,P_q'\rangle_{\rm conn} \simeq \sum_{g=0}^{\infty}e^{-(2g-2+n)S_0}O_{g,p,q}(P_1,...,P_p;P_1',...,P_q')\,,
	\eeq
	where $O_{0,1,1}(P;P') = \delta(P-P')$, reproducing our genus-0 inner product.  More generally,
	\beq
		O_{g,p,q}(P_1,...,P_p;P_1',...,P_q') = (-1)^{3g-3+n}V_{g,n}(2\pi i \sqrt{P_1} ,...,2\pi i \sqrt{P_q'} ) \,.
	\eeq
	\item We then evolve to the asymptotic future using future de Sitter trumpets, with $\widehat{V} |P_1,...,P_p\rangle = | p_1 = P_1,...,p_p=P_p\rangle$.
\end{enumerate}
We represent this interpretation in Fig.~\ref{F:dSquantum}.

The connected contributions with only future universes can be understood as giving the connected part of the Hartle-Hawking wavefunction,
\beq
	\langle p_1,...,p_p|\text{HH}\rangle_{\rm conn} \simeq \sum_{g=0}^{\infty} e^{-(2g-2+p)S_0} \Psi_{g,p}(p_1,...,p_p)\,,
\eeq
where $\Psi_{0,1}(p) = \delta'(p-1)$, $\Psi_{0,2}(p_1,p_2) = \Theta(p_1)\delta(p_1-p_2)$, and more generally
\beq
	\Psi_{g,p}(p_1,...,p_p) = (-1)^{3g-3+p} V_{g,p}(2\pi i \sqrt{p_1} ,...,2\pi i \sqrt{p_p})\Theta(p_1) \cdots \Theta(p_p)\,.
\eeq
The Hartle-Hawking state arises from semi-infinite evolution of the no-boundary state as 
\beq
	|\text{HH}\rangle = \widehat{V} |\varnothing\rangle\,.
\eeq

The end result is a simple picture for the infinite-time evolution operator appearing in the dS JT $S$-matrix. It is
\beq
	\widehat{\mathcal{U}} = \widehat{V} \widehat{\mathcal{O}}\widehat{V}^{\dagger} + |\text{HH}\rangle\langle \text{HH}| + \cdots = \widehat{V}\left( \widehat{\mathcal{O}}+|\varnothing\rangle\langle \varnothing| + \cdots\right)\widehat{V}^{\dagger}
\eeq
where the $\cdots$ correspond to other components with disconnected contributions between the past and future.  Given that the Hartle-Hawking state is non-normalizable, it is natural to project it out under evolution, leading to the simpler $\widehat{\mathcal{U}}' = \widehat{V}\widehat{\mathcal{O}}\widehat{V}^{\dagger}$.

At leading order in the genus expansion, this latter evolution operator $\widehat{\mathcal{U}}'$ acts unitarily, as the identity, on the subspace of states spanned by $|p_1',...,p_q'\rangle$ with $p_i'>0$. However the genus corrections on that subspace are Hermitian and so this emergent unitarity is only approximate and not preserved by the genus expansion.

\subsubsection{Topological recursion and holographic dual}

The amplitudes of Euclidean AdS JT gravity are known to obey topological recursion, and furthermore can be matched to the genus expansion of a certain formal matrix integral. Because dS JT amplitudes are their suitable continuation, it is natural to expect them to also obey topological recursion and to perhaps even follow from a certain matrix integral. Here we show the former, and explain the sense in which the latter is true.

The dS JT amplitudes follow from a continuation of AdS amplitudes whereby $S_0 \to S_0 + i\frac{3 \pi}{2}$ and we continue from positive $\text{Re}(\beta)$ to negative $\text{Re}(\beta)$. The former immediately and strongly suggests that dS JT gravity is formally dual to the analytic continuation of the AdS JT matrix integral under $e^{2S_0} \to - e^{2S_0}$. In a finite $N$ matrix model the number of degrees of freedom is $N^2$. In the double scaling limit $e^{2S_0}$ is the effective number of degrees of freedom. So under this continuation we have $N_{\rm eff} \to - N_{\rm eff}$. 

Let us presume a matrix model interpretation for dS JT amplitudes with a Hamiltonian $\widetilde{H}$ and density of states $\widetilde{\rho}$, along with the dictionary that the $n$-point matrix model connected average of $\text{tr}\!\left( e^{-\beta \widetilde{H}}\right)$ computes the $n$-boundary connected amplitude,
\beq
	\left\langle \text{tr}\!\left( e^{-\beta_1\widetilde{H}}\right)\cdots\text{tr}\!\left( e^{-\beta_n \widetilde{H}}\right)\right\rangle_{\rm MM,conn} \simeq \sum_{g=0}^{\infty} e^{-(2g-2+n)S_0} \widetilde{Z}_{g,n}(\beta_1,...,\beta_n)\,.
\eeq
Writing $\left\langle \text{tr}\!\left( e^{-\beta \widetilde{H}}\right)\right\rangle_{\rm MM}  = \int dE \,\langle \widetilde{\rho}(E)\rangle_{\rm MM} e^{-\beta E}$ with $\langle \widetilde{\rho}(E)\rangle_{\rm MM}$ the ``average density of states'' we extract its genus-0 approximation from the disk amplitude~\eqref{E:dSdisk} via
\beq
	\widetilde{Z}_{0,1}(\beta) = \frac{1}{\sqrt{2\pi}(-\beta)^{3/2}}e^{\frac{1}{\beta}} = \int_{-\infty}^0 dE \,\widetilde{\rho}_0(E) e^{-\beta E}\,,
\eeq
with 
\beq
	\widetilde{\rho}_0(E) = \frac{1}{\sqrt{2}\pi} \sin(2\sqrt{-E})\,,
\eeq
or equivalently
\beq
	\langle \widetilde{\rho}(E)\rangle \approx \frac{e^{S_0}}{\sqrt{2}\pi}\sin(2\sqrt{-E})\Theta(-E)\,.
\eeq
There are two striking features of this quantity. The first is that it cannot be interpreted as an actual density of states since it oscillates from positive to negative. The second is that it is valued on the negative real axis, instead of the positive real axis as one finds in ordinary JT gravity. This is commensurate with $\text{Re}(\beta)<0$.

A similar computation shows that the cylinder amplitude gives the connected two-point function of $\widetilde{\rho}(E)$ to be
\beq
\label{E:twopoint0}
	\langle \widetilde{\rho}(E)\widetilde{\rho}(E')\rangle_{\rm MM,\,conn} \approx \frac{1}{(2\pi)^2}\frac{E+E'}{\sqrt{-E}\sqrt{-E'}(E-E')^2}\,\Theta(-E)\Theta(-E')\,.
\eeq
Even though $\widetilde{\rho}(E)$ cannot be interpreted as a density of states, the expression~\eqref{E:twopoint0} is the usual one for long-range level repulsion in double-scaled matrix models, upon accounting for the fact that the ``cut'' runs along the negative real axis. As a result the genus-0 ``double resolvent''
\beq
	\widetilde{R}_{0,2}(E,E') = \int_{-\infty}^0 d\lambda d\lambda' \langle \widetilde{\rho}(\lambda)\widetilde{\rho}(\lambda')\rangle_{\text{MM,\,conn,\,}g=0} \,\frac{1}{E-\lambda}\frac{1}{E'-\lambda'}
\eeq
takes the form appropriate for a double-scaled matrix integral with a cut along $E<0$.

From the $n$-point function of the ``density of states'' we also have
\beq
	\left\langle \text{tr}\left( \frac{1}{E_1-\widetilde{H}}\right) \cdots \text{tr}\left( \frac{1}{E_n-\widetilde{H}}\right)\right\rangle_{\rm MM,conn} \simeq \sum_{g=0}^{\infty} e^{-(2g-2+n)}\widetilde{R}_{g,n}(E_1,...,E_n)\,,
\eeq 
where for $E>0$ we have
\beq
	\widetilde{R}_{g,n}(E_1,...,E_n) = \int_{-\infty}^0 d\beta_1 \cdots d\beta_n \,e^{\beta_1E_1 + \cdots + \beta_n E_n} \widetilde{Z}_{g,n}(\beta_1,...,\beta_n)\,.
\eeq

From the leading ``density of states'' we define an elliptic curve as one ordinary does in double scaled matrix integrals, namely by analytically continuing $\pi i \widetilde{\rho}_0(E)$ just below the cut to complex values and then defining $\widetilde{z}^2 = E$.  The curve is then
\beq
	\widetilde{y}(\widetilde{z}) =  -\frac{1}{\sqrt{2}}\sinh(2z)\,.
\eeq
We then define
\beq
	\widetilde{W}_{g,n}(\widetilde{z}_1,...,\widetilde{z}_n) = (-2)^n \widetilde{z}_1 \cdots \widetilde{z}_n \widetilde{R}_{g,n}(\widetilde{z}_1^2,...,\widetilde{z}_n^2)\,,
\eeq
along with $\widetilde{W}_{0,1}(\widetilde{z}) = -2 \widetilde{z}\widetilde{y}(\widetilde{z})$ and $\widetilde{W}_{0,2}(\widetilde{z}_1,\widetilde{z}_2) = \frac{1}{(\widetilde{z}_1-\widetilde{z}_2)^2}$. Apart from $g=0$ and $n=1,2$, the $\widetilde{W}_{g,n}$'s are related to the Weil-Petersson volumes in the following way. By~\eqref{E:masterAmplitude} we have
\begin{align}
	\widetilde{W}_{g,n}(\widetilde{z}_1,...,\widetilde{z}_n) = & (-2)^n \widetilde{z}_1 \cdots \widetilde{z}_n \int_{-\infty}^0 d\beta_1 \cdots d\beta_n \,e^{\beta_1 \widetilde{z}_1^2 + \cdots + \beta_n \widetilde{z}_n^2} 
	\\
	\nonumber
	& \times (-1)^{3g-3+n} \int_0^{\infty} d\alpha_1^2 \cdots d\alpha_n^2 \,V_{g,n}(2\pi i \alpha_1,...,2\pi i \alpha_n) \widetilde{Z}_T(\beta_1,i\alpha_1) \cdots \widetilde{Z}_T(\beta_n,i \alpha_n)\,,
\end{align}
where $\widetilde{Z}_T(\beta,i \alpha) = \frac{1}{\sqrt{2\pi}(-\beta)^{1/2}}e^{\frac{\alpha^2}{\beta}}$. Using
\beq
	\int_{-\infty}^0 d\beta\,e^{\beta \widetilde{z}^2} \frac{1}{\sqrt{2\pi}(-\beta)^{1/2}}e^{\frac{\alpha^2}{\beta}} = \frac{e^{-2\widetilde{z}\alpha}}{\sqrt{2}\widetilde{z}}\,,
\eeq
we then have
\beq
\label{E:dSW}
	\widetilde{W}_{g,n}(\widetilde{z}_1,...,\widetilde{z}_n) = 2^{n/2} (-1)^{3g-3} \int_0^{\infty} d\alpha_1^2 \cdots d\alpha_n^2 \,e^{-2\widetilde{z}_1\alpha_1 - \cdots - 2\widetilde{z}_n\alpha_n} V_{g,n}(2\pi i \alpha_1,...,2\pi i \alpha_n) \,.
\eeq

In ordinary AdS JT gravity with the same conventions the genus-0 density of states and elliptic curve are
\beq
	\rho_0(E) = \frac{1}{\sqrt{2}\pi}\sinh(2\sqrt{E})\Theta(E)\,, \qquad y(z) = \frac{1}{\sqrt{2}}\sin(2z)\,, 
\eeq
with $z^2 = -E$. The connected resolvent expansion coefficients are for $E<0$
\beq
	R_{g,n}(E_1,...,E_n) = (-1)^n \int_0^{\infty} d\beta_1 \cdots d\beta_n^2 \,e^{\beta_1 E_1 + \cdots + \beta_n E_n} Z_{g,n}(\beta_1,...,\beta_n)\,,
\eeq
and we define
\beq
	W_{g,n}(z_1,...,z_n) = (-2)^n z_1 \cdots z_n R_{g,n}(-z_1^2,...,-z_n^2)\,,
\eeq
as well as $W_{0,1}(z) = 2z y(z)$ and $W_{0,2}(z_1,z_2) = \frac{1}{(z_1-z_2)^2}$. Using the form of the AdS amplitudes in~\eqref{E:AdSamplitudes} we have apart from those special case
\beq
	W_{g,n}(z_1,...,z_n) = 2^{n/2} \int_0^{\infty} db_1^2 \cdots db_n^2 \,e^{-2 z_1 b_1 - \cdots - 2 z_n b_n} V_{g,n}(2\pi b_1,...,2\pi b_n)\,.
\eeq
The $W_{g,n}$'s are known to obey topological recursion using the spectral curve $y(z)$~\cite{Eynard:2007kz,Saad:2019lba}. 

By a contour rotation argument the integral in~\eqref{E:dSW} is
\begin{align}
\begin{split}
	\int_0^{\infty} d\alpha_1^2 \cdots &d\alpha_n^2 \,e^{-2\widetilde{z}_1\alpha_1 - \cdots - 2\widetilde{z}_n\alpha_n} V_{g,n}(2\pi i \alpha_1,...,2\pi i \alpha_n)  
	\\
	&= (-1)^n \int_0^{\infty} db_1^2 \cdots db_n^2 \,e^{-2z_1 b_1 - \cdots - 2z_n b_n}V_{g,n}(2\pi b_1,...,2\pi b_n)\,,
\end{split}
\end{align}
where $z_j = i \widetilde{z}_j$. The above shows that apart from $g=0$ and $n=1,2$ the $W_{g,n}$'s are related to the $\widetilde{W}_{g,n}$'s as
\beq
\label{E:Weq0}
	\widetilde{W}_{g,n}(\widetilde{z}_1,...,\widetilde{z}_n) = (-1)^{3g-3+n} W_{g,n}(i \widetilde{z}_1,...,i \widetilde{z}_n)\,.
\eeq
In fact, owing to
\beq
	\widetilde{y}(\widetilde{z}) = i y(i \widetilde{z})
\eeq
and
\beq
	\widetilde{W}_{0,2}(\widetilde{z}_1,\widetilde{z}_2) = \frac{1}{(\widetilde{z}_1-\widetilde{z}_2)^2} =- W_{0,2}(z_1 = i \widetilde{z}_1,z_2 = i \widetilde{z}_2)\,,
\eeq
we find that~\eqref{E:Weq0} holds for the special cases $g=0$ and $n=1,2$ as well. The punchline is that the dS JT spectral curve $\widetilde{y}(\widetilde{z})$ and coefficients $\widetilde{W}_{g,n}$'s are simple analytic continuations of their AdS cousins. 

So defined, the $\widetilde{W}_{g,n}$'s obey topological recursion with the spectral curve $\widetilde{y}(\widetilde{z})$, essentially because the $W_{g,n}$ obey topological recursion with $y(z)$ (in the conventions of~\cite{Saad:2019lba} our $W_{g,n}$'s are those with $\alpha = \gamma = \frac{1}{2\pi^2}$). This gives further evidence that dS JT gravity is dual to the analytic continuation of the ordinary JT matrix integral.

\subsection{Topological gauge theory}
\label{S:BF}

So far our approach has been to start with the measure $\text{Pf}(-\Omega)$ along with $\text{Re}(\beta)<0$ and to evaluate the sum over $(-,-)$ $R=2$ metrics with this measure.  In particular, we saw that this prescription is consistent with a positive-definite inner product of asymptotic states in de Sitter.  Let us take a step back and show that this prescription follows from a suitable sum over metrics.  Such a sum is easiest to carry out in the first order formalism, along the lines of~\cite{Saad:2019lba, Cotler:2019nbi} (see also~\cite{Iliesiu:2019xuh}).

Specifically, we can recast dS JT gravity, interpreted as a sum over $(-,-)$ metrics, as a $\text{PSL}(2;\mathbb{R})$ topological gauge theory. First we package the zweibein and spin connection as an $\mathfrak{sl}(2;\mathbb{R})$-valued 1-form.  Consider the generators
\begin{align}
P_0 = \frac{1}{2}\begin{pmatrix}
1 & 0 \\ 0 & -1
\end{pmatrix}\,, \qquad P_1 = \frac{1}{2} \begin{pmatrix}
0 & 1 \\ -1 & 0
\end{pmatrix}\,, \qquad \Omega = \frac{1}{2} \begin{pmatrix}
0 & 1 \\ 1 & 0
\end{pmatrix}\,,
\end{align}
which form a fundamental representation of $\mathfrak{sl}(2;\mathbb{R})$ satisfying
\begin{align}
[P_a, P_b] = \epsilon_{ab}\,\Omega\,,\qquad [\Omega,P_a] = \epsilon_{ab} P^b\,.
\end{align}
Letting $e^a$ for $a=1,2$ be the zweibein and $\omega = - \frac{1}{2} \epsilon^{ab}\omega_{ab}$ be the abelian spin connection, we construct
\begin{align}
A = e^2 P_0 + \omega\, P_1 - e^1  \Omega\,.
\end{align} 
We further introduce an $\mathfrak{sl}(2;\mathbb{R})$-valued scalar $B$ defined by
\begin{align}
B = \frac{1}{4}\left(t^0 P_0 + t^1 P_1 + \varphi\, \Omega\right)
\end{align}
where $t^0, t^1$ are Lagrange multiplier fields which will ultimately impose the torsion-free constraint, and $\varphi$ is the dilaton. In terms of these the first-order dS JT action can be written as
\begin{align}
\label{E:BF1}
S_{\text{JT}} &= S_0 \chi + \int_M d^2 x \, \sqrt{-g}\,(\phi (R - 2) + 2 t^a \mathcal{T}_a )  - 2 \int_{\partial M}dx\, \sqrt{\gamma}\phi(K-1)\nonumber \\
&= S_0 \chi + \int_M \text{tr}(B F) - 2 \int_{\partial M}  dx\,\text{tr}(B A_x)
\end{align}
where $F$ is the field strength of $A$ and $\mathcal{T}^0, \mathcal{T}^1$ are the components of the torsion.  The action of~\eqref{E:BF1} is that of a BF theory, where $B$ enforces $F = 0$, localizing solutions to the symplectic moduli space of flat connections.
Let us carefully treat the action~\eqref{E:BF1} at the quantum level, including the appropriate apparatus of gauge-fixing auxiliary fields and ghosts along the lines of~\cite{Witten:1991we}.

To establish notation, let $\star$ be the Hodge star, and let $T$ flip of the sign of the negative component of the metric on the Lie algebra.  Then we define the operator $\hat{\star} := \star \,T$.  With this notation, let us show how the measure over the symplectic moduli space of flat connections arises in the BF language.  For ease of explanation, it suffices to consider the BF path integral $Z_{\Sigma}$ on a closed orientable 2-manifold $\Sigma$.  Letting $A_{(0)}$ be a flat connection, we write $A = A_{(0)} + \bar{A}$ and let $D_{(0)}$ denote the gauge-covariant differential at $A_{(0)}$.  To compute the partition function we fix the gauge
\begin{align}
\label{E:gaugefixing1}
\hat{\star} \,D^{(0)}\hat{\star}\,\bar{A} = 0\,.
\end{align} 
Then the partition function is given by
\begin{align}
Z_\Sigma = \frac{1}{\# Z(\text{PSL}(2;\mathbb{R}))}\int [dA]\,[dB]\,[dc]\,[d\bar{c}]\,[dw]\,e^{i \int_\Sigma \text{tr}(BF) - i \int_\Sigma \text{tr}(w(D^{(0)} \hat{\star} \bar{A})) - \int_\Sigma \text{tr}(D^{(0)}\bar{c} \,\wedge\, \hat{\star} D^{(0)} c)}\,.
\end{align}
Above, $\# Z(\text{PSL}(2;\mathbb{R})) = 1$ is the size of the center of $\text{PSL}(2;\mathbb{R})$, $w$ is a Lagrange multiplier imposing the gauge-fixing condition, and $c, \bar{c}$ are the Faddeev-Poppov ghosts.  To treat the measure factors, we require an inner product on field fluctuations which we regard as a Riemannian metric on field space.  Since our gauge field has a non-compact gauge group, it is sensible to choose~\cite{Bar-Natan:1991fix, Saad:2019lba} 
\begin{align}
\label{E:gaugemetric1}
G_A( \delta A_1 , \delta A_2)  &:= \frac{1}{\pi^2} \int_\Sigma \text{tr}(\delta A_1 \wedge \hat{\star} \,\delta A_2) = \frac{1}{\pi^2} \int_\Sigma d^2 x \,\sqrt{g}\, g^{\mu \nu}\text{tr}(\delta A_{1,\mu}\cdot T\delta A_{2,\nu}) \,.
\end{align}
We are working in a convention in which $\sqrt{g} > 0$ for a $(-,-)$ signature metric.  The inclusion of $T$ in $\hat{\star}$ is usually designed to render $G_A$ positive-definite~\cite{Bar-Natan:1991fix, Saad:2019lba}, but in our setting $G_A$ is negative-definite since the metrics $g_{\mu \nu}$ constructed out of $A$ are in $(-,-)$ signature.  Then the measure $[dA] = \sqrt{G_A}\, [dA]_{\text{flat}}$ is the square root determinant of the metric times the flat measure.  Since $G_A$ is negative definite, the square root determinant $\sqrt{G_A}$ contributes a factor of $i$ for each mode.

Turning next to the bosonic scalars $B,w$ and the fermionic scalars $c,\bar{c}$, we observe that so long as we choose the same type of inner product for all of these fields, the corresponding square root determinant terms will cancel out~\cite{Witten:1991we}.  We will choose instead to give the same type of inner product for the bosons and fermions up to signs.  As such, we make the convenient choice
\begin{align}
\begin{split}
G_B( \delta B_1 , \delta B_2 ) &:= \pm \frac{1}{\pi^2} \int_\Sigma \text{tr}(\delta B_1 \wedge \hat{\star} \,\delta B_2)\,,
\\
G_w( \delta w_1 , \delta w_2 ) &:= \pm \frac{1}{\pi^2} \int_\Sigma \text{tr}(\delta w_1 \wedge \hat{\star} \,\delta w_2)\,, \\
G_{c,\bar{c}}(\delta \bar{c}_1 , \delta c_2) &:= \mp \frac{1}{\pi^2} \int_\Sigma \text{tr}(\delta \bar{c}_1 \wedge \hat{\star} \,\delta c_2)\,,
\end{split}
\end{align}
where we have allowed for the possibility two distinct sign choices.  Either way, the signs for the $G_B$ and $G_w$ are the same, and are opposite that of $G_{c,\bar{c}}$.  We can think of $G_B$ and $G_w$ as covariant metrics, and $G_{c,\bar{c}}$ as a covariant skew-symmetric form.  Let $M$ be a shorthand for the number of modes of $B$ (which is infinite).  Each of $w$, $c$ and $\bar{c}$ likewise have $M$ modes.  Computing the associated volume forms of $G_B, G_w, G_{c,\bar{c}}$\,, with either sign choice we find the path integral measure
\begin{align}
\begin{split}
\label{E:volumeforms1}
[dB]\,[dw]\,[dc]\,[d\bar{c}] &= (-1)^M\frac{\sqrt{G_B} \sqrt{G_w}}{G_{c,\bar{c}}}\,[dB]_{\text{flat}}\,[dw]_{\text{flat}}\,[dc]_{\text{flat}}\,[d\bar{c}]_{\text{flat}} \\
&= (-1)^M [dB]_{\text{flat}}\,[dw]_{\text{flat}}\,[dc]_{\text{flat}}\,[d\bar{c}]_{\text{flat}} \,.
\end{split}
\end{align}

With this measure in hand the integral over $B$ and $w$ is
\begin{align}
	Z_{\Sigma} = \int [dA]\,[dc]\,[d\bar{c}]\,e^{- \int_\Sigma D^{(0)} \bar{c} \, \wedge \, \hat{\star} D^{(0)} c} \delta[F] \, \delta[\hat{\star} \,D^{(0)} \hat{\star} \,\bar{A}]\,,
\end{align}
and the delta functionals impose
\begin{align}
\label{E:localeq1}
0 = F = D^{(0)} \hat{\star}\,\bar{A}\,.
\end{align}
These are local equations which carve out the moduli space of flat connections $\mathscr{M}$ in the space of all connections.  (Note that the gauge-fixing condition is important for this `carving out'; if we did not impose the gauge-fixing condition, then we would have a space bigger than the true $\mathscr{M}$.)  Since we have assumed that $A_{(0)}$ is flat, then $\bar{A} = 0$ is a solution to~\eqref{E:localeq1}.  But it might not be an isolated solution.

Integrating out $c$ and $\bar{c}$ gives
\begin{align}
\label{E:cintout1}
	Z_{\Sigma} = \int [dA]\, \delta[F]\, \delta[\hat{\star} \,D^{(0)} \hat{\star}\,\bar{A}]\,\det \Delta_0[A_{(0)}]\,,
\end{align}
where $\Delta_0 = \hat{\star} \,D^{(0)} \hat{\star} \,D^{(0)}$ acts on zero forms (and above we have notated its dependence on $A_{(0)}$.)  Using $D^{(0)} \bar{A} = d\bar{A} + A_{(0)} \wedge \bar{A} + \bar{A} \wedge A_{(0)}$,~\eqref{E:cintout1} can be written as
\begin{align}
\label{E:partition4}
	Z_{\Sigma} = \int [dA]\, \delta[D^{(0)} \bar{A} + \bar{A} \wedge \bar{A}]\, \delta[\hat{\star} \,D^{(0)} \hat{\star} \,\bar{A}]\,\det \Delta_0[A_{(0)}]\,,
\end{align}
and the constraints are now
\begin{align}
\label{E:constraints2}
0 = D^{(0)} \bar{A} + \bar{A} \wedge \bar{A} = D^{(0)} \hat{\star}\, \bar{A}\,.
\end{align}
Following~\cite{Witten:1991we} we parse out the remaining analysis into two cases; the first is less realistic, but easier to grasp.

\subsubsection*{First case: isolated solutions}

For simplicity, suppose that the \textit{only} solution to the constraints is $\bar{A} = 0$, so that $A_{(0)}$ is an isolated solution.  We will relax this (unrealistic) assumption shortly.  In this setting, we can replace $[dA]$ in~\eqref{E:partition4} with $[d\bar{A}]$ since $A_{(0)}$ is fixed.  Linearizing~\eqref{E:constraints2} around $\bar{A} = 0$, we obtain
\begin{align}
(D  \oplus \hat{\star} \,D\,\hat{\star}) \,\delta \bar{A} = 0\,,
\end{align}
which we have written in slightly more compact notation.  We have also dropped the $(0)$ superscript on $D$ since the context is clear.  Writing $Q = D \oplus \hat{\star} \,D\,\hat{\star}$, we evaluate~\eqref{E:partition4} to be
\begin{align}
\label{E:partition5}
	Z_{\Sigma} = \frac{\det \Delta_0}{|\det Q\,|}\,.
\end{align}
The absence of the $(-1)^M$ factor can be understood as follows.  Recall from earlier that $[dA]$ carries an $i$ for each mode and thus so does $[d\bar{A}]$.  In the present setting, the number of modes of $\bar{A}$ is the same as the number of modes of $B$ and $w$ jointly, namely $2M$, since these are the Lagrange multiplier fields producing the constraint which exactly exhaust all modes of $\bar{A}$.  Thus $[d\bar{A}] = i^{2M}\, [d\bar{A}]_{\text{flat}} = (-1)^{M}\, [d\bar{A}]_{\text{flat}}$, and so this $(-1)^M$ cancels out the factor in~\eqref{E:volumeforms1}.

Let us evaluate $|\det Q\,| = \sqrt{\det Q Q^\dagger}$ in~\eqref{E:partition5}.  Recall that $Q = D \oplus \hat{\star} \,D\,\hat{\star}$ where $Q : \Omega^1 \to \Omega^2 \oplus \Omega^0$.  Then $Q^\dagger = D \oplus \hat{\star} \,D\,\hat{\star}$ where $Q^\dagger : \Omega^0 \oplus \Omega^2 \to \Omega^1$.  (Note that the daggering flips both the arrow, as well as the order of $\Omega^0$ and $\Omega^2$).  Let $\omega = \omega_0 \oplus \omega_2$ be a sum of a 0-form and a 2-form.  Then
\begin{align}
Q^\dagger \omega = D \omega_0 + \hat{\star} \,D \,\hat{\star}\, \omega_2\,,
\end{align}
Applying $Q$, we find
\begin{align}
\begin{split}
Q Q^\dagger \omega &= \hat{\star} \,D \,\hat{\star} \,D \omega_0 \,\oplus\,\left( \hat{\star} \,D\,\hat{\star}\,\hat{\star}\,D\,\hat{\star} + D\,\hat{\star}\,D\,\hat{\star} \right) \omega_2  \\
&= \hat{\star} \,D\,\hat{\star}\,D \omega_0 \,\oplus\,\left( D \,\hat{\star} \,D\,\hat{\star} \right) \omega_2  \\
&= (\Delta_0 \oplus \Delta_2) \,\omega\,,
\end{split}
\end{align}
and so $Q Q^\dagger = \Delta_0 \oplus \Delta_2$.  Here we observe that, on 2-forms, $(\hat{\star} \, D \,\hat{\star} \,D + D \,\hat{\star} \,D \,\hat{\star}) \omega_2 = D \,\hat{\star}\, D \,\hat{\star}\, \omega_2$, since $\hat{\star} \,D \,\hat{\star} \,D \omega_2 = 0$ (because $D \omega_2 = 0$). Thus $|\det Q\,| = \sqrt{\det \Delta_0 \, \det \Delta_2}$.

With our expression for $|\det Q\,|$ at hand~\eqref{E:partition5} simplifies to
\begin{align}
\label{E:partition6}
Z_{\Sigma} = \sqrt{\frac{\det \Delta_0}{\det \Delta_2}}\,.
\end{align}
On an orientable manifold $\Sigma$ $\hat{\star}$ is a well-defined isomorphism.  Suppose $v$ is a 0-form eigenfunction of $\Delta_0$, so that $\Delta_0 f = \lambda \, f$.  But then $(\hat{\star} \,\Delta_0 \,\hat{\star}) (\hat{\star} f) = \lambda (\hat{\star} f)$, and so $\Delta_2 (\hat{\star} f) = \lambda (\hat{\star} f)$.  We can similarly run the argument the other direction.  As such, there is a 1-to-1 identification between the eigenfunctions and eigenvalues of $\Delta_0$ and $\Delta_2$, so that $\det \Delta_0 = \det \Delta_2$.  Then~\eqref{E:partition6} becomes, simply, $1$.  We can think of this as
\begin{align}
Z_{\Sigma} = \text{vol}(\mathscr{M})
\end{align} 
for $\text{vol}(\mathscr{M}) = 1$.  The result should be viewed as the contribution of the moduli space in the (trivial) connected component of our fixed flat connection $A_{(0)}$.

\subsubsection*{Second case: non-isolated solutions}

Suppose, for a given flat connection $A_{(0)}$, that the set of solutions to~\eqref{E:constraints2} is not isolated.  We also assume that $\mathscr{M}$ forms a manifold with a single connected component, but it is simple to generalize our analysis to multiple connected components.  These conditions imply that, for a particular $A_{(0)}$, there are infinitesimal fluctuations $\delta \bar{A}$ which satisfy~\eqref{E:constraints2} to linear order to $\delta \bar{A}$.  Then the tangent space to the space of flat connections at $A_{(0)}$ can be written as
\begin{align}
T_{A_{(0)}} \mathscr{M} \oplus \Omega_{\perp}^1
\end{align}
where $\Omega_{\perp}^1$ is just defined as the orthocomplement of $T_{A_{(0)}} \mathscr{M}$.  So fluctuations that move us along the moduli space correspond to $\delta \bar{A}$ solely in $T_{A_{(0)}} \mathscr{M}$.  (In the previous case, this tangent space was trivial.)  The symplectic form and the inner products restrict to ones on $T_{A_{(0)}} \mathscr{M}$ and $\Omega_{\perp}^1$.

At each fixed flat connection $A_{(0)}$, let us construct $\delta \bar{A}_{\perp}[A_{(0)}]$ which live in $\Omega_{\perp}^1$.  We will just write $\delta \bar{A}_{\perp}[A_{(0)}]$ as $\delta \bar{A}_\perp$, where the dependence on $A_{(0)}$ is understood.  Then~\eqref{E:partition4} can be written as
\begin{align}
\label{E:partition8}
Z_{\Sigma} = \int_{\mathscr{M}} [dA_{(0)}] \, \det \Delta_0\,\int_{\Omega_\perp^1} [d\delta \bar{A}_\perp]\,\delta[D^{(0)} \delta \bar{A}_\perp]\,\delta[\hat{\star}\,D^{(0)} \,\hat{\star}\,\delta \bar{A}_\perp]\,.
\end{align}
Here we are justified in only integrating over linearized perpendicular fluctuations $\delta \bar{A}_\perp$, since path integrals over delta functionals are 1-loop exact.  Letting $Q_{\perp}[A_{(0)}] := Q|_{\Omega_\perp^1}$, the $\delta \bar{A}_\perp$ integral evaluates to $1/|\det Q_\perp|$ and so~\eqref{E:partition8} becomes
\begin{align}
\label{E:partition9}
Z_{\Sigma} = \int_{\mathscr{M}} [dA_{(0)}] \, \frac{\det \Delta_0}{|\det Q_\perp|}\,.
\end{align}
There is no $(-1)^M$ for the same reason as in the isolated case: there are as many modes of $\delta \bar{A}_\perp$ as there are of $B$ and $w$ jointly, and so $[d\bar{A}_\perp] = i^{2M} [d\bar{A}_\perp]_{\text{flat}} = (-1)^M [d\bar{A}_\perp]_{\text{flat}}$.  Then this $(-1)^M$ cancels out the one in~\eqref{E:volumeforms1}.

Going forward we will assume that $\Delta_0$ has no kernel for simplicity, as occurs for $\Sigma$ with genus greater than $1$.  Now what does $Q_\perp$ actually mean?  Examining~\eqref{E:partition8}, we see that $D^{(0)} \delta \bar{A} = 0 = \hat{\star}\,D^{(0)}\,\hat{\star} \,\delta \bar{A} = 0$ means that $\delta \bar{A}$ is both closed and co-closed with respect to $D^{(0)}$, and hence harmonic.  Thus, $\delta \bar{A} \in T_{A_{(0)}}\mathscr{M}$ is the same as saying that $\delta \bar{A}$ is harmonic.  Accordingly, $\Omega_\perp^1$ is the orthocomplement of harmonic 1-forms.  Let $P_\perp$ be a projection onto $\Omega_\perp^1$.  Then we should replace $Q$ with $Q P_\perp$ and $Q^\dagger$ with $P_\perp Q^\dagger$.  Then $QQ^\dagger$ is replaced with $Q P_\perp Q^\dagger$ in our analysis.

Recalling that $Q^\dagger \omega = D \omega_0 + \hat{\star}\,D\,\hat{\star} \,\omega_2$, applying $P_\perp$ means that we project out harmonic forms.  But being harmonic is equivalent to being both closed and co-closed.  Note that closed here means $D Q^\dagger \omega = D \,\hat{\star}\, D \,\hat{\star} \,\omega_2 = 0$, and co-closed here means $\hat{\star} \,D \,\hat{\star}\, Q^\dagger \omega = \hat{\star} \,D\,\hat{\star}\, D \omega_0 = 0$.  Equivalently, projecting out harmonic forms means projecting out $\omega_0$ which satisfy $\Delta_0 \omega_0 = 0$, and projecting out $\omega_2$ which satisfy $\Delta_2 \omega_2 = 0$.  This is the same as omitting the zero modes of $\Delta_0$ and $\Delta_2$.  In total, we find
\begin{align}
| \det Q_\perp| = \sqrt{\det Q P_\perp Q^\dagger} = \sqrt{\det \widetilde{\Delta}_0 \, \det \widetilde{\Delta}_2}
\end{align}
where the tilde denotes the omission of zero modes.  But since the spectra of $\Delta_0$ and $\Delta_2$ are the same,~\eqref{E:partition9} becomes
\begin{equation}
\label{E:A0int0}
Z_{\Sigma} = \int_{\mathscr{M}} [dA_{(0)}] = (-1)^{\dim(\mathscr{M})/2}\, \text{vol}(\mathscr{M})
\end{equation}
on account of the remaining factors of $i$ in $[dA_{(0)}]$.  Thus we land on the claimed result
\begin{align}
Z_\Sigma = \int_{\mathscr{M}} \text{Pf}(-\Omega)\,.
\end{align}
That is, the factor of $(-1)^{\dim(\mathscr{M})/2}$ in~\eqref{E:A0int0} (which originally comes from the negative-definiteness of $G_A$ in~\eqref{E:gaugemetric1}) explains where the minus sign in $\text{Pf}(-\Omega)$ comes from.  A similar analysis holds in the more general context of manifolds $\Sigma$ with boundaries, likewise giving rise to $\text{Pf}(-\Omega)$.

Since our argument for obtaining $\text{Pf}(-\Omega)$ relies upon the negative-definite metric $G_A$ in~\eqref{E:gaugemetric1}, it is instructive to compute the metric explicitly in a special case to gain intuition.  To this end, we evaluate the induced metric on field space for fluctuations in the double trumpet for Euclidean AdS in $(-,-)$ signature.  In this case we can write
\begin{align}
e^1 = b \cosh(dx + f'(\rho)\,d\rho)\,,\quad e^2 = d\rho\,,\quad \omega = b \sinh(\rho)\,(dx + f'(\rho) \rho)\,,
\end{align}
where here $f(\rho)$ encodes the twist $\tau\sim \tau+2\pi$ by interpolating between $f(-\infty) = 0$ and $f(\infty) = \tau$.  We can build up an $A$ from $e^1, e^2, \omega$, and consider fluctuations around this background.  A gauge fluctuation of $b$ can be written as
\begin{align}
\delta_b A = \frac{db}{2}\begin{pmatrix}
- \frac{1}{b}\,\text{sech}^2(\rho)\,d\rho & e^{-\rho}(1 + \tanh(\rho))\,dx \\ \\
e^\rho (1 - \tanh(\rho))\,dx & \frac{1}{b}\,\text{sech}^2(\rho)\,d\rho
\end{pmatrix}
\end{align}
which satisfies the requisite gauge-fixing condition~\eqref{E:gaugefixing1}.  Then we compute 
\begin{align}
G_A(\delta_b A, \delta_b A) =\frac{1}{\pi^2} \int d^2 x \, \sqrt{g}\,g^{\mu \nu} \text{tr}(\delta_b A_\mu \cdot T \delta_b A_\nu) = - \frac{ db^2}{b}\,.
\end{align}
We can similarly compute $G_A(\delta_\tau A, \delta_\tau A)$ for a twist fluctuation $\delta_\tau A$ by the following trick.  Since we know from the symplectic form $\Omega(\delta_1 A, \delta_2 A) = \frac{1}{\pi^2} \int \text{tr}\left( \delta A_1 \wedge \delta A_2\right)$ that the joint measure over $b$ and $\tau$ is $\frac{b\,db\,d\tau}{\pi} $, we have
\begin{align}
\sqrt{\left(-G_A(\delta_b A, \delta_b A) \right)\left(-G_A(\delta_\tau A, \delta_\tau A) \right)} = \frac{b\,db\,d\tau}{\pi} \,,
\end{align}
from which we infer
\begin{align}
G_A(\delta_\tau A, \delta_\tau A) = - \frac{b^3\,d\tau^2}{\pi^2}\,.
\end{align}
Then the induced metric on $b,\tau$ fluctuations descending from $G_A$ is thus
\begin{align}
ds^2 = - \left(\frac{db^2}{b}  +  \frac{b^3 \, d\tau^2}{\pi^2}\right)\,,
\end{align}
which is negative-definite as expected.

\subsection{Summary}

There are a number of major results derived in this Section which we summarize here for the reader's benefit. 

In Subsection~\ref{S:BF} we show how to arrive at the symplectic measure $\text{Pf}(-\Omega)$ on the moduli space of constant curvature metrics in $(-,-)$ signature from a sum over metrics in the first-order formalism. That measure, together with a $90^{\circ}$ rotation of the field contours of the Schwarzian modes mandated by the $i\epsilon$ prescription of dS JT gravity ($\text{Re}(\beta)<0$), led to the desired topological expansion.

These choices have practical benefits. The global dS$_2$ amplitude, presented in~\eqref{E:globalAmplitude}, can be computed from a sum over Lorentzian $R=2$ metrics. That sum includes a moduli space integral that converges only with the $i\epsilon$ prescription above. Crucially, the inner product of asymptotic states obtained in~\eqref{E:innerproduct1} is positive-definite only thanks to these choices.

The amplitude $\widetilde{Z}_{g,n}$ of dS JT gravity on a general genus $g$ surface with $n$ boundaries can be written as~\eqref{E:masterAmplitude2}. For the disk see instead~\eqref{E:dSdisk} and the cylinder~\eqref{E:doubleTrumpet}. While we arrived at these amplitudes directly from a sum over hyperbolic metrics in $(-,-)$ signature, we found in Subsection~\ref{S:AdStodS} that they can also be obtained from a suitable continuation from AdS JT amplitudes. That continuation involves two steps:
\begin{enumerate}
	\item An imaginary additive renormalization $S_0\to S_0 + i\frac{3\pi}{2}$, so that the effective string coupling $g_s = e^{-S_0} \to i e^{S_0}$ becomes pure imaginary.
	\item A suitable continuation from $\text{Re}(\beta)>0$ to $\text{Re}(\beta)<0$.
\end{enumerate}
The first implies that dS JT gravity has a holographic dual of a sort, namely the formal analytic continuation of the AdS JT matrix integral under $e^{2S_0} \to -e^{2S_0}$. The resulting object is not itself a matrix integral; the ``density of states'' extracted from gravity amplitudes is not strictly positive. Nevertheless the dS amplitudes, suitably massaged, do obey topological recursion.

We also presented a bulk quantum mechanical interpretation of the genus expansion in Subsection~\ref{S:dSQM}. The final result is pictured in Fig.~\ref{F:dSquantum} and described in the surrounding text.

\section{Doubly non-perturbative effects}

In the last Section we established the genus expansion of dS JT gravity. In AdS JT gravity that expansion is asymptotic, and relatedly, there are contributions to amplitudes which are non-perturbatively suppressed in the effective string coupling $e^{-S_0}$. Thinking of the contribution of different topologies as already being non-perturbative, these effects are said to be doubly non-perturbative. 

The goal of this Section is to identify some simple examples of such effects in dS JT gravity.

\subsection{Borel-Le Roy resummation}

As we have seen, the dS JT path integral measure contributes alternating minus signs in the genus expansion.  These signs render certain quantities Borel--Le Roy resummable~\cite{le2012large}.  Let us give two examples.  First, consider the no-boundary partition function, which receives contributions from all spacetimes without boundary. This quantity diverges on account of the divergent sphere amplitude but its higher-genus corrections are finite. These are captured by the sum 
\begin{align}
\sum_{g=2}^\infty e^{(2-2g)S_0}\widetilde{Z}_{g,0} \simeq \sum_{g=2}^\infty (-1)^g e^{(2-2g) S_0} \frac{(4\pi^2)^{2g - \frac{5}{2}}}{2^{1/2} \pi^{3/2}}\, \Gamma\!\left(2g - \frac{5}{2}\right)\,,
\end{align}
where we have considered an expansion for large $g$~\cite{Saad:2019lba}.  Due to the alternating signs, one can perform a Borel--Le Roy resummation without encountering a Borel pole on the positive real axis in the Borel plane.

As another example, consider higher-genus corrections to the Hartle-Hawking wavefunction of the no-boundary state with a single future boundary. In JT gravity this comes from a sum over surfaces with no past boundaries and a single future boundary. In the $p$-basis, that wavefunction reads
\begin{equation}
\langle p| \text{HH} \rangle \!\simeq\!   e^{S_0} \delta'(p-1) + \sum_{g=1}^\infty (-1)^{g}e^{(1-2g) S_0} V_{g,1}(2\pi i p) \Theta(p)\,.
\end{equation}
For $g \gg p$ and $g \gg 1$, the Weil-Petersson volumes go as~\cite{Saad:2019lba} (accounting for the fact that we are working in the convention that the normalization $\alpha$ of the symplectic form appearing in~\cite{Saad:2019lba} is $\frac{1}{2\pi^2}$)
\begin{equation}
V_{g,1}(2\pi i p) \approx \frac{1}{2\pi}\left( \frac{2}{\pi^2}\right)^g\, \Gamma\!\left(2g - \frac{3}{2}\right)\,\frac{\sin\!\left(\sqrt{\frac{\pi p}{2}}\right)}{ \sqrt{p}}\,.
\end{equation}
At fixed $p$, on account of the alternating minus signs, there is again no obstruction to Borel--Le Roy resummation.

We emphasize that this resummation is contingent on a certain order of limits: we compute amplitudes at fixed $p_i$'s and topology, and then sum over topologies.  One could instead attempt to sum over the $p$'s first, i.e.~to compute the amplitude in the $\Phi$-basis, and then sum over genera. 

These topological expansions are resummable and their resummation gives a candidate doubly non-perturbative definition of dS JT amplitudes. However, this is not the whole story. There is an ambiguity in the non-perturbative completion of the genus expansion since instead we could define dS JT gravity through its formal holographic dual, the analytic continuation of the AdS JT matrix integral under $e^{2S_0} \to -e^{2S_0}$. Under that definition there are further eigenvalue instanton corrections to amplitudes. Even then, this holographic dual almost certainly has ambiguities since the original model dual to AdS JT gravity has doubly non-perturbative ambiguities~\cite{Saad:2019lba, Johnson:2019eik, Eynard:2023qdr}.

\subsection{A dS version of topological gravity}
\label{S:dSAiry}

In Euclidean AdS JT gravity,  single- and multiple-eigenvalue instantons in the dual formal matrix integral contribute doubly non-perturbative effects to the $S$-matrix, for instance corrections to the density of states that are doubly non-perturbatively suppressed in $S_0$. Noting that the genus expansion of the density of states is supported on $E \geq 0$, the instantons provide contributions which are oscillatory $\sim e^{i e^{S_0}}$ for $E \geq 0$ and exponentially decay as $e^{-e^{S_0}}$ for $E \leq 0$. Under the assumption that these effects contribute to dS JT gravity as well under the analytic continuation $S_0\to S_0 + i \frac{3\pi}{2}$ we note that this behavior is exchanged, with instanton corrections being a sum of doubly non-perturbatively suppressed $\sim e^{-e^{S_0}}$ and growing behavior $\sim e^{e^{S_0}}$ for $E>0$  and oscillatory behavior $\sim e^{ie^{S_0}}$ in the $E<0$ region. These are potentially enormous effects which could in principle drastically change the (meta-)observables of dS JT gravity.

However there is some reason to think this is not the case. Recall that our dS model can be thought of as arising from the combination of the analytic continuation $S_0 \to S_0 + i\frac{3\pi}{2}$ along with a continuation in $\beta$'s. The latter amounted to the statement that the dS amplitudes, upon Laplace transforming to functions of energy in a putative dual, become functions along the negative real axis. A useful prototype to keep in mind is the double-scaling limit of the Airy model, dual to topological gravity, under the same continuation $S_0\to S_0 + i\frac{3\pi}{2}$ as in the map from AdS to dS JT gravity. We consider this to be a de Sitter version of topological gravity, which we explore further in Appendix~\ref{App:Airy1}.

The double-scaling limit of the Airy model mostly depends on the coupling $S_0$ and energy through a variable $\xi = - e^{\frac{2S_0}{3}}E$ so that the exact, non-perturbative density of states can be written as
\begin{equation}
\label{E:toflip1}
\langle \rho(E) \rangle = e^{\frac{2S_0}{3}} \left(\text{Ai}'(\xi)^2 - \xi \text{Ai}(\xi)^2\right)\,.
\end{equation}
Under our analytic continuation we find in Appendix~\ref{App:Airy1} that the exact density of states is mapped as $\langle \rho(E)\rangle \to \langle \rho(-E)\rangle$. In particular, the leading term in the genus expansion is
\begin{equation}
\langle \rho(E)\rangle_{g = 0} =\frac{e^{S_0}}{\pi}\sqrt{-E}\,\Theta(-E) + O(e^{-S_0})\,,
\end{equation}
and the full non-perturbative result is
\begin{equation}
\langle \rho(E) \rangle = e^{\frac{2S_0}{3}} \left(\text{Ai}'(-\xi)^2 + \xi \text{Ai}(-\xi)^2\right)\,.
\end{equation}
Notice that the doubly non-perturbative contributions just serve to flip~\eqref{E:toflip1} from the positive to the negative real axis.  As such, the dS Airy model does have a microscopic dual: we can construct a finite-dimensional GUE matrix with a cut along the negative axis which approaches the dS Airy model in the double-scaling limit. Furthermore, because the cut now runs along the negative axis, the natural objects to probe the model with are insertions of $\text{tr}\!\left( e^{-\beta H}\right)$ with $\beta<0$, just as one does in dS JT gravity.

The doubly nonperturbative story for dS JT is less clear.  In that setting, the leading density of states is $\frac{e^{S_0}}{\sqrt{2}\pi}\,\sin(2 \sqrt{-E})\,\Theta(-E)$. This too is supported along negative energies, but it has the strange feature of oscillating negative on account of the sine and so cannot be interpreted as a physical density of states. That is, to the extent that the dual exists, it is not a matrix integral but only a formal analytic continuation thereof. However, near the edge we indeed recover the dS Airy leading density of states $\propto e^{S_0}\,\sqrt{-E}\,\Theta(-E)$.  One takeaway is that the positivity of the dS Airy density of states is somewhat of an accident. If we consider the continuation of other $(2,2p-1)$ double scaled matrix models dual to minimal models coupled to gravity, giving ``de Sitter versions'' thereof, one ends up with a leading ``density of states'' that is non-negative near $E=0$, but which for $p>1$ has regions where it is negative.

\section{Discussion}

This work completes a long saga establishing the foundations of dS JT gravity.  Some crucial ingredient of this work include (i) the definition of a topological expansion in which we sum over $(-,-)$ signature hyperbolic metrics, and (ii) a modification of the path integral measure in conjunction with an $i\epsilon$ prescription.  Taken together, we find that dS JT has a pure imaginary effective string coupling and so an alternating-sign genus expansion.

dS JT gravity then provides the first truly non-perturbative understanding of dS quantum gravity and holography.  In particular, dS JT gravity can be defined non-perturbatively as a suitable continuation of the AdS version and formally has a holographic dual continuation of the AdS JT matrix integral.  We suspect that some of the lessons learned here about the path integral measure will generalize to higher dimensions.  For instance, in 3+1 dimensions, subtleties with the de Sitter path integral measure may be related to old questions about the phase of the $\mathbb{S}^4$ partition function~\cite{Polchinski:1988ua, Anninos:2020hfj}.  We are currently investigating this possibility~\cite{CotlerWIP}.

JT gravity can be viewed as a certain $p \to \infty$ limit of the $(2,2p-1)$ minimal string~\cite{Mertens:2020hbs}. Versions of the $(2,2p-1)$ minimal string describing de Sitter spacetimes have been previously explored in~\cite{Martinec:2003ka}, although the results were inconclusive.  Our findings in the present paper, combined with those in~\cite{Cotler:2023eza}, may help us properly treat the de Sitter $(2,2p-1)$ minimal string; as such the latter merits revisiting.  It would also be interesting to see if there is a de Sitter version of the Virasoro minimal string~\cite{Collier:2023cyw}.  In that setting, to access de Sitter physics one might want to couple two Liouville theories with central charges $c = 13 \pm i \lambda$ for $\lambda$ large.

In some circumstances, one can view dS JT gravity as a description of the near-horizon of near-Nariai black holes in 3+1 dimensions~\cite{Maldacena:2019cbz}.  It would be interesting to see if some of our results appear in that setting.  This being said, we emphasize that our treatment of dS JT gravity in the present paper does not presuppose any connection to near-Nariai black holes, and instead treats dS JT gravity on its own terms.

More broadly, it is intriguing that dS JT gravity is dual to a formal matrix integral under the continuation $N_{\rm eff} \to - N_{\rm eff}$.  This continuation resembles the putative duality between dS Vasiliev gravity in 3+1 dimensions and the $\text{Sp}(N)$ model of anticommuting scalars in three dimensions.  Recall that Euclidean AdS Vasiliev gravity in four dimensions is thought to be dual to the singlet sector of the $O(N)$ model with (commuting) scalars in three dimensions (see e.g.~\cite{Giombi:2016ejx} for a review).  The amplitudes of the dS model are obtained from those of the AdS model by a similar continuation $N \to -N$, which in CFT amounts to mapping the $O(N)$ commuting scalars to $\text{Sp}(N)$ anticommuting scalars~\cite{Anninos:2011ui}. In particular, there is a sense in which $O(-N)$ is $Sp(N)$ at the level of representation theory.  Since for a vector model $N_{\text{eff}} = N$, its continuation is similarly $N_{\rm eff} \to - N_{\rm eff}$.  It is tempting to surmise that an effectively negative number of degrees of freedom is a general feature of de Sitter holography.

It is not clear how to formulate finite-dimensional matrix models  with the property that $N_{\text{eff}}$ is negative, and which limit to the dS JT model in a double-scaling limit.  A natural guess is to consider fermionic matrix models, such as adjoint fermionic matrix models~\cite{Makeenko:1993jg, Ambjorn:1994bp, Semenoff:1996vm, Paniak:2000zy}.  For such models, one can consider even-powered matrix potentials $\sum_{n \geq 0}  c_{2n}\,\text{tr}(\bar{\Psi} \Psi)^n$ where $\bar{\Psi} \Psi$ is a fermion matrix bilocal.  While fermion loops in matrix fat graphs indeed contribute factors of $-1$, the matrix model genus expansion is complicated by the fact that lines must go from $\Psi$ to $\bar{\Psi}$.  While there are known cases in which the free energy of an adjoint fermion matrix model has alternating signs in the genus expansion~\cite{Makeenko:1993jg, Ambjorn:1994bp, Semenoff:1996vm}, it is not clear how to construct adjoint fermionic matrix models which reproduce the genus expansions of e.g.~the GUE or JT gravity but with alternating signs.  If such a construction is possible, it would seem to require matrix potentials of the form $\sum_{n \geq 0}  c_{n}\,\text{tr}(\bar{\Psi} \Psi)^{n/2}$, since $\sqrt{\bar{\Psi} \Psi}$ often acts as a stand-in for a Hermitian matrix $H$.  It is not clear at a technical level how to treat half-powers of the fermion bilocal.  We note that for the adjoint fermion matrix model with potential $\bar{\Psi} \Psi$, it appears that a natural resolvent to compute is $\text{tr}\!\left(\frac{1}{z - \sqrt{\bar{\Psi} \Psi}}\right)$ whereas in the literature $\text{tr}\!\left(\frac{1}{z - \bar{\Psi} \Psi}\right)$ is computed instead~\cite{Makeenko:1993jg, Semenoff:1996vm}.

Technical issues aside, it is not clear if we should expect there to be a fermionic manifestation of dS JT in the first place.  For example, even in the putative duality between de Sitter Vaseliev gravity and $\text{Sp}(N)$ spinless fermions, it appears that the connection to fermions only holds for a 3+1-dimensional bulk.  In other dimensions there is an analytic continuation of $N$ but not an accompanying fermionic interpretation.  This and other features of de Sitter Vasiliev gravity will be explored in~\cite{CotlerWIP2}.

Going forward, we aim to extend our understanding of dS JT to higher dimensions, which holds significant promise. Pure 2+1-dimensional de Sitter Einstein gravity is a natural first target~\cite{Cotler:2019nbi}.  Then perhaps we can ultimately apply similar methods to 3+1-dimensional de Sitter quantum gravity, apropos to our own universe.

\subsection*{Acknowledgements}
We thank H.~Chen, B.~Eynard, A.~Kar, J.~Kruthoff, J.~Maldacena, D.~Marolf, F.~Rosso, and J.~Turiaci for valuable discussions. JC is supported by the Harvard Society of Fellows and the Simons Collaboration on Celestial Holography.  KJ thanks the Kavli Institute for Theoretical Physics for their hospitality while this work was being completed, and was supported in part by NSERC and by the National Science Foundation under Grant No. NSF PHY-1748958.

\appendix

\section{Density of states in the de Sitter Airy model}
\label{App:Airy1}

Here we derive the non-perturbative density of states in the ordinary Airy model and its de Sitter counterpart.  We begin by recalling some more general structural facts about finite-dimensional matrix models, and then specialize to the Airy setting.
\subsection{Orthogonal polynomial preliminaries}

Consider a finite degree polynomial potential $V(\lambda)$ such that $\int d\lambda \, e^{-V(\lambda)} < \infty$.  Then we can uniquely define the real orthogonal polynomials $\{P_n(\lambda)\}_{n \geq 0}$ by the following conditions:
\begin{enumerate}
\item $P_n(\lambda)$ is degree $n$ and has the form $P_n(\lambda) = \lambda^n + \cdots$.
\item $\int d\lambda\,P_m(\lambda)\,P_n(\lambda)\,e^{- V(\lambda)} = h_m \, \delta_{mn}$ where $h_m > 0$.
\end{enumerate}
Now consider an $N \times N$ Hermitian matrix model with probability density $\mathcal{P}(H) \propto e^{- \text{tr}(V(H))}$.  This induces a probability density over the $N$ eigenvalues given by (see e.g.~\cite{ginsparg1993lectures})
\begin{align}
\label{E:P1}
\mathcal{P}(\lambda_1,...,\lambda_N) = \frac{1}{Z_N}\, \prod_{i < j} (\lambda_i - \lambda_j)^2\, e^{- \sum_{i=1}^N V(\lambda_i)}\,,
\end{align}
where $Z_N$ is a constant which we will compute shortly.

There is a useful identity which allows us to write~\eqref{E:P1} in terms of the orthogonal polynomials $\{P_n(\lambda)\}_{n \geq 0}$.  Letting $[P_{j-1}(\lambda_i)]_{i,j=1,...,N}$ denote the $N \times N$ matrix whose $(i,j)$ entry is $P_{j-1}(\lambda_i)$, we have the identity~\cite{ginsparg1993lectures}
\begin{align}
\prod_{i < j} (\lambda_i - \lambda_j) = \det\, [P_{j-1}(\lambda_i)]_{i,j=1,...,N}\,.
\end{align}
Accordingly,~\eqref{E:P1} can be rewritten as
\begin{align}
\label{E:P2}
\mathcal{P}(\lambda_1,...,\lambda_N) = \frac{1}{Z_N} \Big(\det \,[P_{j-1}(\lambda_i)]_{i,j=1,...,N}\Big)^2\, e^{- \sum_{i=1}^N V(\lambda_i)}\,.
\end{align}

First let us compute the normalization $Z_N$ using the data of the orthogonal polynomials.  Since $\mathcal{P}(\lambda_1,...,\lambda_N)$ is a probability density it satisfies $\int d\lambda_1 \cdots d\lambda_N \, \mathcal{P}(\lambda_1,...,\lambda_N) = 1$.  Using~\eqref{E:P2} and expanding out the determinant, we find
\begin{align}
\int d\lambda_1 \cdots d\lambda_N \, \mathcal{P}(\lambda_1,...,\lambda_N)  &= \frac{1}{Z_N}\int d\lambda_1 \cdots d\lambda_N \left(\sum_{\sigma \in S_N}\!\!\left(\text{sgn}(\sigma)\,\prod_{j=1}^N P_{\sigma(j)-1}(\lambda_{j})\right)\right)^{\!\! 2} e^{- \sum_{i=1}^N V(\lambda_i)}\,.
\end{align}
But using the orthogonality properties of the polynomials, the only terms which do not integrate to zero are those for which all $P_{j-1}(\lambda_i)$'s are paired with another $P_{j-1}(\lambda_i)$.  Thus the above integral reduces to
\begin{align}
\frac{1}{Z_N} \sum_{\sigma \in S_N} \prod_{j=1}^N \left(\int d\lambda_j \,P_{\sigma(j)-1}(\lambda_j)^2\, e^{- V(\lambda_j)}\right) = \frac{N! \, \prod_{j=1}^N h_{j-1} }{Z_N}\,.
\end{align}
Since the above must equal one, we obtain
\begin{align}
Z_N = \frac{1}{N! \prod_{j=1}^N h_{j-1}}\,.
\end{align}

We can use a similar approach to compute the density of states.  Notice from~\eqref{E:P1} that $\mathcal{P}(\lambda_1,...,\lambda_N)$ is symmetric, and as such the exact density of states $\langle \rho(\lambda) \rangle$ is given by e.g.
\begin{align}
\langle \rho(\lambda) \rangle = N \int d\lambda_2 \cdots d\lambda_N\, \mathcal{P}(\lambda, \lambda_2,...,\lambda_N)\,.
\end{align}
By a similar computation as the one we did above to compute the normalization, we find
\begin{align}
\label{E:rhoexact1}
\langle \rho(\lambda) \rangle = \sum_{j=1}^N \frac{1}{h_{j-1}}\,P_{j-1}(\lambda)^2\, e^{- V(\lambda)}\,.
\end{align}
The above can be written in a nice way upon introducing some suggestive notation.  Let us define
\begin{align}
\psi_n(\lambda) := \frac{1}{\sqrt{h_n}}\,P_{n}(\lambda)\, e^{-\frac{1}{2}\,V(\lambda)}\,.
\end{align}
Then~\eqref{E:rhoexact1} becomes
\begin{align}
\label{E:rhoexact2}
\langle \rho(\lambda) \rangle = \sum_{n=0}^{N-1} \psi_n(\lambda)^2\,.
\end{align}
Sometimes the $\psi_n(\lambda)^2$ above is written as $|\psi_n(\lambda)|^2$ since $\psi_n(\lambda)$ is usually real, but we will avoid this notation to prevent confusion when we consider the de Sitter setting later on.

There is a nice way to simplify~\eqref{E:rhoexact2}; for this we consider the slightly more general kernel
\begin{align}
K(\lambda_1, \lambda_2) = \sum_{n=0}^{N-1} \psi_n(\lambda_1)\,\psi_n(\lambda_2)\,,
\end{align}
which reduces to~\eqref{E:rhoexact2} when $\lambda_1 = \lambda_2 = \lambda$.  Now the Christoffel-Darboux formula for orthogonal polynomials gives us
\begin{align}
K(\lambda_1, \lambda_2) = \sqrt{\frac{h_{N}}{h_{N-1}}} \, \frac{\psi_{N}(\lambda_1) \psi_{N-1}(\lambda_2) - \psi_{N}(\lambda_2) \psi_{N-1}(\lambda_1)}{\lambda_1-\lambda_2}\,.
\end{align}
Taking $\lambda_1, \lambda_2 \to \lambda$, the above becomes
\begin{align}
\label{E:rhoexact3}
\langle \rho(\lambda) \rangle = K(\lambda, \lambda) = \sqrt{\frac{h_{N}}{h_{N-1}}} \left(\psi_{N}'(\lambda) \psi_{N-1}(\lambda) - \psi_{N}(\lambda) \psi_{N-1}'(\lambda)\right)
\end{align}
We will use this formula extensively in the next Subsection.

\subsection{Exact Airy density of states}

With the above technology at hand, we now turn to the Airy model.  We start with a Gaussian Hermitian matrix model, and performing an appropriate double scaling limit.  Let the initial potential be $V(\lambda) = \frac{2N}{a^2}\,\lambda^2$.  In a large $N$ expansion, the genus zero density of states of the Hermitian matrix model with this potential is the famous Wigner semicircle law
\begin{align}
\label{E:Wigner1}
\langle \rho(\lambda)\rangle_0 = \frac{e^{S_0}}{\pi} \sqrt{\frac{a^2 - \lambda^2}{2a}}\,\Theta(a^2 - \lambda^2)\,, \qquad e^{S_0} := \frac{N}{\left(\frac{a}{2}\right)^{3/2}}\,.
\end{align} 
The support of the density of states is $-a \leq \lambda \leq a$.  In the present setting the orthogonal polynomials $\{P_n(\lambda)\}_{n \geq 0}$ are Hermite polynomials, 
\begin{align}
P_n(\lambda) = \left(\frac{a^2}{8N}\right)^{\! n/2} H_n\!\left(\frac{\sqrt{2N}}{a}\,\lambda\right)\,,
\end{align}
and the normalization coefficients $h_n$ are
\begin{align}
h_n = \frac{n!}{2^n}\,.
\end{align}
It is easy to see that the $\psi_n(\lambda)$'s are $L^2$-normalized quantum harmonic oscillator wavefunctions, satisfying
\begin{align}
\label{E:Sch1}
\left(- \frac{d^2}{d\lambda^2} + \left(\frac{2N}{a^2}\right)^{\! 2} \lambda^2\right)\! \psi_n(\lambda) = \frac{4N}{a^2}(n + 1/2)\,\psi_n(\lambda)\,.
\end{align}
Defining the creation operator
\begin{align}
\mathcal{D} := \frac{\sqrt{N}}{a}\left(x - \frac{a^2}{2N} \frac{d}{dx}\right),
\end{align}
we have standard the relation
\begin{align}
\mathcal{D} \psi_n(\lambda) = \sqrt{n+1}\, \psi_{n+1}(\lambda)\,.
\end{align}
With these notations in mind, we can rewrite~\eqref{E:rhoexact3} in this setting as
\begin{align}
\label{E:rhoexact4}
\langle \rho(\lambda) \rangle = \frac{\sqrt{N}}{2} \left(\mathcal{D}\psi_{N-1}'(\lambda) \, \psi_{N-1}(\lambda) - \mathcal{D}\psi_{N-1}(\lambda) \, \psi_{N-1}'(\lambda)\right)\,.
\end{align}

To take the double-scaling limit we zoom in on the left edge of the density of states. To do so we take $\lambda \to \lambda - a$ so that the large $N$, genus zero density of states is supported on $0 \leq \lambda \leq 2a$, 
\begin{align}
\label{E:Wigner2}
\langle \rho(\lambda)\rangle_0 = \frac{e^{S_0}}{\pi} \sqrt{\lambda - \frac{\lambda^2}{2a}}\,\Theta\!\left(\lambda - \frac{\lambda^2}{2a}\right)\,, \qquad e^{S_0} := \frac{N}{\left(\frac{a}{2}\right)^{3/2}}\,.
\end{align} 
Then we simultaneously take $a \to \infty$ and $N \to \infty$ in such a way that the genus zero density of states stays finite; examining~\eqref{E:Wigner2}, we see that we should keep the ratio $e^{S_0}$ fixed.  In this so-called double-scaling limit, we find
\begin{align}
\label{E:Wigner3}
\langle \rho(\lambda)\rangle_0 = \frac{e^{S_0}}{\pi} \sqrt{\lambda}\ \Theta(\lambda)\,.
\end{align}
Turning to~\eqref{E:Sch1}, let us shift $\lambda \to \lambda - a$ and define $\Psi(x) := \psi_{N-1}(\lambda - a)$.  Substituting $N = e^{S_0} \left(\frac{a}{2}\right)^{3/2}$, the differential equation for $\Psi(x)$ in the double scaling limit becomes
\begin{align}
\label{E:Sch2}
\left(- e^{-2 S_0} \frac{d^2}{d\lambda^2} - \lambda\right) \Psi(\lambda) = 0\,.
\end{align}
Above we have multiplied through by a factor of $e^{- 2 S_0}$ so that $e^{-S_0}$ plays the role of $\hbar$.  The solution to the above equation is the Airy function, namely
\begin{align}
\Psi(\lambda) \propto \text{Ai}(\xi)\,, \qquad \xi := - e^{\frac{2S_0}{3}} \lambda\,.
\end{align}
What we have really learned is that in the double scaling limit, $\psi_{N-1}(\lambda - a) \sim C(a, e^{S_0}) \, \text{Ai}(\xi)$ where $C(a, e^{S_0})$ is some function of $a$ and $e^{S_0}$.  We will not need the explicit form of this function to proceed.  Observe that the double-scaling limit of~\eqref{E:rhoexact4} (i.e.~first shifting $\lambda \to \lambda - a$ and then taking the double-scaling limit) is
\begin{align}
\begin{split}
\langle \rho(\lambda) \rangle &\propto \text{Ai}'(\xi)^2 - \text{Ai}''(\xi) \text{Ai}(\xi) \\
&\propto \text{Ai}'(\xi)^2 - \xi\,\text{Ai}(\xi)^2\,,
\end{split}
\end{align}
where in going from the first line to the second line we have used~\eqref{E:Sch2} to simplify $\text{Ai}''(\xi)$.  To fix the constant of the proportionality for $\langle \rho(\lambda) \rangle$, we can expand in large $e^{S_0}$ to recover the genus zero term~\eqref{E:Wigner3}; this shows us that
\begin{align}
\label{E:nonpertDOS1}
\langle \rho(\lambda) \rangle = e^{\frac{2S_0}{3}}\left(\text{Ai}'(\xi)^2 - \xi\,\text{Ai}(\xi)^2\right)\,.
\end{align}
which is the full, non-perturbative density of states for the Airy model.

\subsection{Exact de Sitter Airy density of states}

Recall that our continuation of the Euclidean AdS JT matrix mode to its de Sitter counterpart has two ingredients: (i) a continuation of $S_0 \to S_0 \pm i \frac{3\pi}{2}$, and (ii) a continuation of temperature $\beta$ so that it has a negative real part.  Let us take as our starting point the Airy model from the previous subsection, and apply the continuation.  In the Airy model, the disk amplitude is
\begin{align}
Z_{0,1}(\beta) = \langle \text{tr}(e^{- \beta H})\rangle_0 = \frac{e^{S_0}}{\sqrt{4\pi} \beta^{3/2}}\,.
\end{align}
Continuing $S_0 \to S_0- i \frac{3\pi}{2}$ and $\beta \to -x < 0$ (through a counterclockwise trajectory around $\beta=0$) as in our map from AdS JT amplitudes to dS JT amplitudes, we arrive at the analytic continuation
\begin{align}
\widetilde{Z}_{0,1}(x) = \frac{e^{S_0}}{\sqrt{4\pi} x^{3/2}}\,,
\end{align}
the disk amplitude of the dS version of the Airy model. This can be written as
\begin{align}
\widetilde{Z}_{0,1}(x) = \int_{-\infty}^0 \!\! d\lambda \,\, e^{x \lambda}\, \frac{e^{S_0}}{\pi}\, \sqrt{-\lambda}\,.
\end{align}
which is the natural version of the Laplace transform since $x < 0$.  Hence we see that the continuation has given us a modified genus zero density of states
\begin{align}
\label{E:WignerdS1}
\langle \rho(\lambda) \rangle_0 = \frac{e^{S_0}}{\pi}\, \sqrt{-\lambda}\, \Theta(-\lambda)\,.
\end{align}

Turning to the exact density of states, let us examine~\eqref{E:Sch2}.  Our continuation of $S_0$ yields
\begin{align}
\label{E:Sch3}
\left(e^{-2 S_0} \frac{d^2}{d\lambda^2} - \lambda\right) \Psi(\lambda) = 0\,.
\end{align}
which has flipped the sign of the kinetic term.  The solutions are
\begin{align}
\Psi(\lambda) \propto \text{Ai}(-\xi)\,, \qquad \xi := - e^{\frac{2S_0}{3}} \lambda\,.
\end{align}
These are related to our previous solutions by $\xi \to - \xi$, which is effectively flipping the sign of the energy.  Repeating the same analysis as in the previous subsection (and fixing the constant of the $\langle \rho(\lambda) \rangle$ by comparing with~\eqref{E:WignerdS1}), we obtain the non-perturbative density of states for the de Sitter version of the Airy model, i.e. its analytic continuation in $S_0$,
\begin{align}
\langle \rho(\lambda) \rangle = e^{\frac{2S_0}{3}}\left(\text{Ai}'(-\xi)^2 + \xi\,\text{Ai}(-\xi)^2\right)\,.
\end{align}
This is simply the original density of states~\eqref{E:nonpertDOS1} under a flip of the energy, $\lambda \to - \lambda$. Equivalently, this is the non-perturbative density of states of an Airy model obtained by zooming in to the \textit{right} square root edge of the spectrum instead of the \textit{left} square root edge.

We also remark that the `Hamiltonian' $\widehat{H}(\lambda) = e^{- 2 S_0} \frac{d^2}{d\lambda^2} - \lambda$ in the original Airy setting of the previous subsection is related to the Hamiltonian in the de Sitter setting by $\widehat{H}(\lambda) \to - \widehat{H}(-\lambda)$.  These signs effectively give a wrong-sign kinetic term in the Hamiltonian.  In the double-scaled Fermi formalism for the ordinary Airy model (see e.g.~\cite{ginsparg1993lectures} for a review), macroscopic loop operator expectation values are computed by certain traces of $e^{- \beta \widehat{H}(\lambda)}$~\cite{banks1990microscopic}.  Apparently in the de Sitter setting, the operator which arises is instead $e^{x (-\widehat{H}(-\lambda))}$ for $x > 0$, which effectively probes thermal distributions of $\widehat{H}(-\lambda)$, which has the right-sign kinetic term.

\section{Comments on the Klein-Gordon inner product}

In this Appendix we consider the Klein-Gordon inner product utilized in~\cite{Maldacena:2019cbz}, and its relation to the inner product used in our paper.  For simplicity we treat a single boundary in the far future (with similar results for a single boundary in the far past).  First let us recall the argument of~\cite{Strobl:1993yn, Maldacena:2019cbz}.  At a large cutoff slice $t = \Lambda$ in the far future, the dS JT boundary conditions are
\begin{align}
ds^2 = - dt^2 + (e^{2\Lambda} + O(1))dx^2\,,\quad \phi = \frac{\Phi}{2\pi}\,e^\Lambda + O(1)\,.
\end{align}
We rewrite the above as
\begin{align}
ds^2 = - dt^2 + \left(\left(\frac{\ell}{2\pi}\right)^2+ O(1)\right)dx^2\,,\quad \phi = \phi_b + O(1)\,,
\end{align}
where the length of the spatial circle is $\ell = 2\pi e^\Lambda$ and the value of the dilaton is $\phi_b = \frac{\Phi}{2\pi}\,e^\Lambda$.  Note that as $\Lambda \to \infty$, we have that $\phi_b /\ell = \Phi/(2\pi)^2$ stays fixed whereas $|\phi_b \ell| \to \infty$.  In the $\phi_b, \ell$ variables, the Wheeler--de Witt equation (i.e.~the Hamiltonian constraint) for a wavefunction $\Psi(\phi_b, \ell)$ is
\begin{align}
\label{E:WdW1}
(\partial_u \partial_v + 1)\Psi = 0\,, \qquad u = \phi_b^2\,,\quad v = \ell^2\,.
\end{align}
We view $\phi_b, \ell$, or alternatively $u,v$, as coordinates on superspace. Since $\phi_b$ and $\ell$ are the only diffeomorphism-invariant quantities on a spatial slice, they parameterize the entire domain of dependence for wavefunctionals.  As such,~\eqref{E:WdW1} is not an approximation (a la minisuperspace) but rather the complete equation for dS JT gravity. For large $|\phi_b \ell|$, the general solution takes the form
\begin{align}
\Psi(\phi_b,\ell) \simeq \frac{1}{2\sqrt{\phi_b \ell}}\, e^{- 2 i \phi_b \ell} f(\phi_b/\ell) + \frac{1}{2\sqrt{\phi_b \ell}}\,e^{2 i \phi_b \ell} \tilde{f}(\phi_b/\ell)\,.
\end{align}
The functions $f, \tilde{f}$ are arbitrary functions which are fixed by boundary conditions outside of the large $|\phi_b \ell|$ regime.  The $e^{\mp 2 i \phi_b \ell}$ factors are due to bulk divergences which are compensated in the path integral by holographic counterterms on the asymptotic boundaries.

Since~\eqref{E:WdW1} takes the form of a Klein-Gordon equation on superspace, there is a conserved $U(1)$ current that we can fashion into an inner product on positive frequency solutions.  That is, we define the inner product
\begin{align}
\label{E:KGIP1}
\langle \Psi_1, \Psi_2\rangle :=  i \int_{-\infty}^\infty \! d\hat{\ell}\,\left(\Psi_1^* \partial_{\hat{\ell}} \Psi_2 - \Psi_2 \partial_{\hat{\ell}} \Psi_1^*\right)
\end{align} 
where $\Psi_1, \Psi_2$ are positive frequency solutions to~\eqref{E:WdW1} (i.e.~they have $\tilde{f} = 0$), and $\hat{\ell} := \frac{\ell}{\phi_b}$.  Note that $\hat{\ell}$ is integrated over the entire real line since $\phi_b$ can take any sign, while the authors of~\cite{Maldacena:2019cbz} only considered positive $\hat{\ell}$. In practice, we evaluate the above integral on a slice in an asymptotically nearly dS$_2$ region where $|\phi_b \ell|$ tends to infinity.

How does~\eqref{E:KGIP1} relate to the inner product we have used in this paper and in our previous work~\cite{Cotler:2019dcj,Cotler:2023eza}?  We propose the following correspondence between the wavefunctional $\Psi(\phi_b,\ell)$ and the dS JT path integral.  Recall that our asymptotic $|\Phi\rangle$-states satisfy
\begin{align}
\langle \Phi | \Phi'\rangle = \sqrt{\Phi} \sqrt{\Phi'}\,\delta(\Phi-\Phi')\,,
\end{align}
and so
\begin{align}
\int_{-\infty}^{\infty} d\Phi\, \frac{|\Phi\rangle \langle \Phi|}{\Phi}  = \mathds{1}\,.
\end{align}
Suppose we have a state $|\psi\rangle$, evolved to future infinity.  Then we propose the correspondence
\begin{align}
\label{E:correspond1}
\Psi(\phi_b, \ell) = \frac{1}{2\sqrt{\phi_b \ell}}\, e^{-2 i \phi_b \ell} \langle \Phi | \psi\rangle \Big|_{\Phi =  (2\pi)^2\,\frac{\phi_b}{\ell}}\,,
\end{align}
where $\langle \Phi|\psi\rangle$ is the transition amplitude computed from the dS JT path integral with final boundary condition $\Phi$ and initial boundary condition corresponding to $|\psi\rangle$. That is, we identify
\begin{align}
f(\phi_b/\ell) = \langle \Phi | \psi\rangle \Big|_{\Phi =  (2\pi)^2\,\frac{\phi_b}{\ell}}\,.
\end{align}
Notice that this identification is the only natural choice; the path integral (with a holographic counterterm) can only produce transition amplitudes depending on $\phi_b/\ell$, and so $f(\phi_b/\ell)$ is the only object with which to identify the transition amplitude. Moreover, near future infinity the path integral only produces positive frequency amplitudes. Constructing $\Psi_1(\phi_b, \ell)$, $\Psi_2(\phi_b, \ell)$ corresponding to $|\psi_1\rangle, |\psi_2\rangle$, at large $|\phi_b \ell|$ we find
\begin{align}
\begin{split}
\langle \Psi_1 , \Psi_2\rangle &= \langle \psi_1| \left(\int_{-\infty}^{\infty} d\Phi\, \frac{|\Phi\rangle \langle \Phi|}{\Phi} \right) |\psi_2\rangle\\
&= \langle \psi_1 | \psi_2\rangle\,.
\end{split}
\end{align}
As such, using~\eqref{E:correspond1}, we have related the Klein-Gordon inner product to ours.

In~\cite{Maldacena:2019cbz}, the authors guessed that~\eqref{E:correspond1} was instead
\begin{align}
\Psi^{\text{them}}(\phi_b, \ell) = \frac{1}{\sqrt{2 \phi_b \ell}}\, e^{-2 i \phi_b \ell} \,\Phi\,\langle \Phi | \psi\rangle \Big|_{\Phi =  (2\pi)^2\,\frac{\phi_b}{\ell}}\,,
\end{align}
or equivalently
\begin{align}
f^{\text{them}}(\phi_b/\ell) = \Phi\,\langle \Phi | \psi\rangle \Big|_{\Phi =  (2\pi)^2\,\frac{\phi_b}{\ell}}\,,
\end{align}
by reasoning about the Hartle-Hawking state (see Eq.~(2.12) of their paper and the surrounding discussion), but this correspondence was not fixed by a calculation.  We propose that~\eqref{E:correspond1} is in fact the proper identification, which leads to an agreement between our inner product and that of~\cite{Maldacena:2019cbz}.

\bibliography{refs}
\bibliographystyle{JHEP}

\end{document}